\documentclass[superscriptaddress,twocolumn]{revtex4-2}
\usepackage{float} 
\usepackage{booktabs}
\usepackage{mathrsfs}
\usepackage{amssymb}
\usepackage{amsmath,empheq}
\usepackage{graphicx}
\usepackage{color}
\usepackage{subfigure}
\usepackage{setspace}
\usepackage[T1]{fontenc}
\usepackage{bm}
\usepackage{upgreek}
\usepackage{orcidlink}
\usepackage{hyperref}
\hypersetup
{
hypertex=true,
colorlinks=true,
linkcolor=blue,
filecolor=blue,
urlcolor=blue,
citecolor=blue,
breaklinks=true
}
\begin{document}

\preprint{APS/123-QED}

\title{Measuring weak microwave signals via current-biased Josephson Junctions II: Arriving at single-photon detection sensitivity
}
\author{Y. Q. Chai\orcidlink{0009-0007-6246-9102}}
\affiliation{HergD collaboration and Information Quantum Technology Laboratory, School of Information Science and Technology, Southwest Jiaotong University, Chengdu 610031, China}
\author{M. Y. Wang\orcidlink{0009-0006-6105-950X}}
\affiliation{HergD collaboration and Information Quantum Technology Laboratory, School of Information Science and Technology, Southwest Jiaotong University, Chengdu 610031, China} 
\author{S. N. Wang\orcidlink{0000-0002-5498-4047}}
\affiliation{HergD collaboration and Information Quantum Technology Laboratory, School of Information Science and Technology, Southwest Jiaotong University, Chengdu 610031, China} 
\author{P. H. Ouyang\orcidlink{0009-0000-5379-8552}}
\affiliation{HergD collaboration and Information Quantum Technology Laboratory, School of Information Science and Technology, Southwest Jiaotong University, Chengdu 610031, China}
\author{L. F. Wei\orcidlink{0000-0003-1533-1550}}
\email{ lfwei@swjtu.edu.cn }
\affiliation{HergD collaboration and Information Quantum Technology Laboratory, School of Information Science and Technology, Southwest Jiaotong University, Chengdu 610031, China}

\date{\today}

\begin{abstract}
It is well known that the current-biased Josephson junction (CBJJ) can serve as a Josephson threshold detector (JTD) for the sensitive detection of weak microwave signals. Based on the recent work (PRB {\bf 111}, 024501 (2025)) on the detection sensitive limit of the usual equilibrium JTD, here we numerically demonstrate that a non-equilibrium JTD can be alternatively utilized to implement the higher sensitive detection of a weak microwave signal, arriving at its energy quantum limit. In the presence of thermal noise, we numerically simulate the phase dynamics for the CBJJ in the JTD with the different sweep rates of the biased currents, and find that the SCDs of the JTD with and without the microwave signal input show different behaviors. It is demonstrated that, depending on how high the sweep rate of the biased current being applied, the JTD can be operated in either the equilibrium- or the non-equilibrium state. Specifically, under the rapidly non-adiabatic driving, the SCDs of the JTD are obviously insensitive to the thermal noises, which means that the non-equilibrium JTD can possess a higher achievable detection sensitivity, compared with its equilibrium state counterpart. Consequently, the non-equilibrium JTD can be utilized to implement the desired single microwave-photon detection. Also, some of the achievable performance indexes, such as the dynamic range, detection bandwidth, and the photon-number resolvability, etc., of the non-equilibrium JTD have been estimated, when it serves as a wideband microwave single-photon detector.
\end{abstract}
\maketitle

%
\section{\label{sec:1}introduction}
It is well known that the sensitive microwave detectors play important roles for various microwave applications, typically in such as wireless communication~\cite{communication}, non-destructive safety detection~\cite{non-destructive}, medical imaging~\cite{imaging}, electronic countermeasures~\cite{electronic1}, and also the electromagnetic responses detection of ultra-high frequency gravitational waves~\cite{gravitational1,gravitational2} and Ultra-light mass dark matter~\cite{axion1,axion2,Staggs_2018,computing}, etc.. However, how to achieve the detection of the microwave signal whose energy approaches, even arrives at its quantum limit (i.e., at the single-photon level) is still a very challenging topic~\cite{ph1,ph2,albertinale_detecting_2021}. The sensitivity of the current microwave receiver with the conventional electromagnetic induction metal antenna is basically limited by the thermal noise, i.e., $10^{-21}~\rm W/\sqrt{Hz}$~\cite{our1} at the room temperature of 290~K. Recently, the graphene-based self-switching diode had demonstrated the weak microwave signal detection with the NEP of about $2.2\times10^{-9}~\rm W/\sqrt{Hz}$~\cite{detector1}, and the Rydberg-atom-based microwave superheterodyne receivers (the sensitivity had realized the microwave electric field detection with the sensitivity of about $55~\rm nV/cm~\sqrt{Hz}$ (The corresponding NEP could be estimated roughly as $9.1\times10^{-18}~\rm W/\sqrt{Hz}$)~\cite{detector3}, etc.. However, the sensitivity of all demonstrated microwave detectors is still far from the desirable single microwave-photon level. Therefore, reducing the operating temperatures of the detectors to suppress the thermal noise significantly is particularly expected. 

Among various cryogenic detectors based on the semiconducting, graphene, and superconducting materials~\cite{semiconductor} and~\cite{graphene,SNSPD,MKID,TES}, devices based on the Josephson junctions (JJs) have demonstrated the great potential to achieve the highly sensitive detection of weak microwave signals~\cite{detector4,detector5,detector6,detector7,detector8,JJph1,JJph2,JJph3,JJph4,JJph5,JJph6,PhysRevApplied.8.024022}. Generally speaking, weak microwave signal detectors based on JJ devices can be divided into two categories: one is by using the JJ-based superconducting-qubit detectors, which realize the desired microwave single-photon detection by probing the superconducting-qubit excitation induced by a single microwave photon. In order to overcome the difficulty of the extremely low excitation probability of a superconducting qubit driven by a single microwave photon, the realization of the superconducting-qubit-based detector of the standing-wave single photon in a microwave cavity has attracted great interest in recent years~\cite{detector6,JJph3,narrowband1,JJph6}. Here, the traveling-wave microwave single photon expected to be detected is transferred first into the standing-wave microwave single photon in the microwave cavity. The injected single photon might modify the qubit excitation frequency in the dispersive doubling regime, and thus could be detected by probing the modified transition frequency of the superconducting-qubit. Although this method could overcome the difficulty of low excitation probability of the superconducting-qubit induced by the traveling-wave microwave photons, the detection efficiency is still limited by the injected probability of the traveling-wave photons, the decoherence duration, and the readout fidelity of the excited superconducting-qubit. 
More importantly, this superconducting-qubit-based microwave single-photon detection approach can only be applied to narrow-band microwave photon detection. Particularly, it is difficult to meet the practical needs for the detection of most microwave signals whose frequencies are usually unknown. Therefore, microwave photon detection based on JJ state switching has received increasing attention~\cite{detector5,graphene_2020,detector4,detector7,JJph1,JJph2,JJph4,JJph5,pankratov_detection_2025}. The basic idea of this method is to detect the induced current caused by the energy absorption of the microwave photons. When microwave photons are absorbed by the JJ, the photon energy is converted into a current of a certain intensity in the JJ. This causes the total current to exceed the critical current of the JJ, thereby breaking the confinement of phase particles in the potential well and causing them to cross the potential barrier, thus switching the JJ from the zero-voltage state to the non-zero-voltage state~\cite{PhysRevB.9.4760}. That is, the switching of the JJ state is regarded as achieving the detection of the incident microwave signal, and this bandwidth microwave photon detector should be signal-frequency insensitive. As the microwave signals with different photon numbers may convert to different amounts of current, this detector naturally possesses the desired photon-number resolution (PNR) ability. 
 
Physically, if the bias current $I_b(t)$ of the JJ is slightly less than its critical current $I_c$, even a very small microwave current $I_s$ may cause the voltage-state switch behavior of the JJ, i.e., it can be switched from the zero-voltage state to the detectable non-zero-voltage one. Specifically, for an underdamped JJ, the switched non-zero-voltage state can persist for a certain time interval, until the JJ returns to its zero-voltage one~\cite{JJph2,JJph4,retrapping,Pankratov2022}. As a consequence, by probing such a stable electrical response voltage signal, the small microwave signal current $I_s$ can be detected effectively. This is the physical basis of the Josephson threshold detector(JTD) for the extremely weak signal detection~\cite{detector4,JJph2,JJph4,JJph6,JJph7,JJPh8}. (Here, we specifically state that the CBJJ can be called a JTD when used to detect weak microwave signals. Therefore, for convenience, in the following, we will describe JJ, CBJJ, and JTD all as JTD.) Of course, due to the influence of various random noises~\cite{noise1,PhysRevLett.93.106801}, the achievable detection sensitivity of the JTD that has been demonstrated in experiments is still quite limited for the weak microwave signal detection~\cite{our2}. This is because, in the practical weak signal detection, the detected switching current $I_{sw}$, when the JTD is switched from the beginning of zero-voltage state to the non-zero-voltage one, actually contains the contribution of various noise currents. Thus, to suppress the influence from the noises, the detection is required to be repeated many times for getting a set of switching current data: $\{I_{sw}^{(n)},\,n=1,2,3,...\}$, where $I_{sw}^{(n)}$ refers to the value of the measured switching current for the $n$-th detection. Obviously, the measured value of each switching current is obtained with a certain probability, relative to the total number of detections. Therefore, the measured switching currents, after a series of repeated detections, show actually a statistical distribution called specifically the switching-current distribution (SCD)~\cite{detector4,JJph2,JJph4,JJph6,JJph7,SW1,SW2,SW3,SW4,SW5,SW6}. 
As a consequence, the detection sensitivity of the JTD can be calibrated by the so-called $d_{KC}$-index~\cite{PhysRevApplied.22.024015,JJph2,JJph4,JJPh8,our1,our2,KC1,PhysRevApplied.16.054015}, which statistically describes the distinction degree between the SCDs for the JTD with and without the application of the detected signal. The equivalent noise power of the detector could be estimated as the signal power corresponding to the minimum $d_{KC}$-index~\cite{our1,our2}. Recently, by the numerical method we demonstrated that~\cite{our1}, the achievable detection sensitivity of the JTD, working in the mK-temperature environment, can approach the energy limit of the applied microwave signal, i.e., at the level of dozen microwave photon energies, if the physical parameters (typically e.g., its the critical current $I_c$) of the used JTD could be optimized effectively. 

Naturally, the next question we want to ask is whether the achievable detection sensitivity of the JTD can realistically arrive at the single-photon level in the microwave band? The answer given by the present work is, Yes, it can. We find that, in almost all of the previous works, attention has been mainly paid to how to optimize the physical parameters, such as inductance, capacitance, and Josephson energy of the used JTD~\cite{our1,our2,detector7,JJph6,JJph7,pankratov_detection_2025}, ignoring the effort to suppress the device's noises. Basically, the phase dynamics of the JTD is nonlinear, and it might be more sensitive to the initial phase of the JTD~\cite{SCD0,nonequilibrium2} than to the thermal noises, under the proper biases. Consequently, if the phase dynamics of the JTD become insensitive to thermal noise, its achievable detection sensitivity for weak signals may be greatly enhanced. The key purpose of the present work is to investigate how such a dynamics can be realized for further improving the achievable detection sensitivity of the JTD to the desired single microwave-photon level.

The remainder of the paper is arranged as follows. By numerically solving the nonlinear phase dynamical equation of the JTD driven by a time-dependent driven current and the usual thermal noise, in Sec.~II, we demonstrate how the initial phase influences the phase dynamics of the JTD under either the ``adiabatic" or ``non-adiabatic" driving~\cite{nonequilibrium1,nonequilibrium2}, and its observable SCD~\cite{SCD0}. This is a basis for the subsequent discussion on how to achieve the detection of a single microwave-photon by using the external magnetic field to modulate the initial phase of the JTD. In Sec.~III, we propose a statistical method to calibrate the detection sensitivity of the JTD by numerically simulating the initial phase dependence of the SCDs, for the JTD with or without the detected microwave signal driving. We show that the ``adiabatic"- and the ``non-adiabatic" driving make the phase of the JTD undergo different dynamics, and thus the JTD can work respectively in the equilibrium- or non-equilibrium states with the different measurable SCDs. Statistically, the Receiver Operating Characteristic (ROC) curves can be used to describe the difference between these SCDs for the JTD with and without microwave signal driving~\cite{ROC1,ROC2,ROC3}. Then, we use the Area Under ROC Curve (AUC) to quantify the successful probability of the detection, i.e., the larger the AUC value, the higher the probability of the detector achieving the weak microwave signal detection. Furthermore, we discuss how to set the initial phase of the JJ for optimizing the detection sensitivity of the JTD. Under the optimal phase condition, we estimate the minimum detectable photon number and the dynamic range of the non-equilibrium JTD. Finally, in Sec.~IV, we summarize our work and discuss the feasibility of the non-equilibrium JTD for the sensitive broadband detection of a single photon in the microwave band.

\section{Phase dynamics for a  driven JTD and its observable voltage-state switches}
For completeness, in this section, we first briefly review the usual phase dynamics for the JTD biased by a slowly varying current. Consequently, the measured SCDs can be well fitted by the relevant semi-empirical formula. Then, we numerically investigate the phase dynamics for the JTD biased by the fast-variable current, yielding the novel SCD with multiple peaks that cannot be fitted by the semi-empirical formula. Its initial phase sensitivity implies that the thermal noise influence could be suppressed effectively. 

Wherein the initial phase is insensitivity, and thus the thermal noise is dominant. As a consequence, the achievable detection sensitivity of such an equilibrium state JTD is limited by the thermal noise. Then, based on the numerical simulations, we demonstrate that the phase dynamics for the JTD, under the ``non-adiabatic" current bias, could be sensitive to the initial phase of the JTD, resulting in the thermal noise being effectively suppressed. Consequently, the JTD can work at the non-equilibrium states and thus possesses the higher sensitivity, over the non-equilibrium JTD, for the microwave photon detection.

\subsection{The usual phase dynamics for a JTD biased by a slowly-variable current: insensitivity to the initial phase of the JTD}

For a biased JTD device, its macroscopic phase dynamics can be generically described by~\cite{SCD0,RCSJ1,RCSJ2,RCSJ3}
\begin{equation}
\begin{aligned}
\frac{{\rm d}^2\varphi}{{\rm d}\tau^2}+\beta\frac{{\rm d}\varphi}{{\rm d}\tau}+\sin(\varphi)=i_b(\tau)+i_n(\tau)\,,
\end{aligned}\label{eq:RCSJ1}
\end{equation}
Here, $\varphi$ is the phase difference between the macroscopic wave functions of the superconductors across the JTD. The noise current $I_n(t)$ and time-dependent bias current $I_b(t)$ are normalized by the JTD critical current $I_c$, yield the dimensionless quantities $i_n(\tau)=I_n(t)/I_c$ and $i_b(\tau)=I_b(t)/I_c$, respectively. Also, $\tau=\omega_Jt$ with $\omega_J=\sqrt{2eI_c/\hbar C}$ being the JTD Plasma frequency. The dimensionless dissipation coefficient of the JTD is given by $\beta=1/(RC\omega_J)$ with $R$ and $C$ being the resistance and capacitance, respectively. Without loss of generality, we assume the time-dependent bias current takes the form of a linearly increasing current, i.e., $i_b(\tau)=v\tau$ (where $v$ is the dimensionless sweep rate of the bias current), used commonly in experiments for measuring the non-zero-voltage state of the JTD~\cite{our2,SCD0}. For the simplicity, the noise current is satisfying the relations~\cite{SCD0,Nyquist,thermal}: $\langle i_n(\tau)\rangle=0, \,\langle i_n(\tau)i_n(\tau')\rangle=2\beta k_BT/E_{J0}$, where $E_{J0}=\hbar I_c/2e$ denotes the Josephson energy of the JTD, $T$ is the working temperature of the JTD and $k_B$ the Boltzmann constant.

\begin{figure}[htbp]
\includegraphics[width=0.8\linewidth]{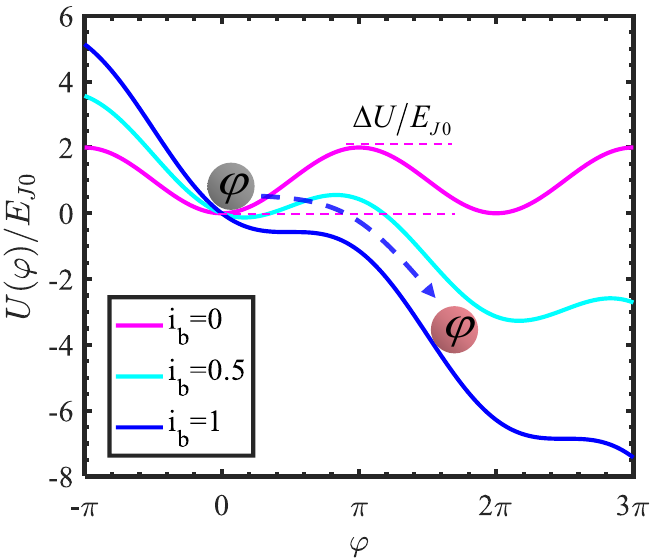}
\caption{A schematic diagram for a phase particle dynamics described by the equation~\eqref{eq:potential}. Here, the gray ball represents the phase particle trapped in a potential well with the height of $\Delta U$, while the pink ball refers to the phase particle that has escaped from the trapped potential well and moves freely.}\label{F1}
\end{figure}

Ideally, when the dissipation and noise effects in the JTD could be neglected, the phase dynamics of the driven JTD reduces to that of a phase particle with an effective mass $m=(\hbar/2e)^{2}C$ moving in a washboard potential field (\ref{eq:potential})
\begin{equation}
\begin{aligned}
U(\varphi)=E_{J0}\left[1-\cos(\varphi)-i_b(\tau)\varphi\right],\,i_b(\tau)<1\,,\label{eq:potential}
\end{aligned}
\end{equation}
shown schematically in Fig.~\ref{F1}. Here, the barrier height of the potential well that traps the phase particle is 
\begin{equation}
\begin{aligned}
\Delta U(\varphi)=2E_{J0}\left[\sqrt{1-i_b(\tau)^2}-i_b(\tau)\arccos(i_b(\tau))\right]\,,
\label{eq:high}
\end{aligned}
\end{equation}
which is the difference between the local minimum and the maximum potential energy of the potential energy $U(\varphi)$. 
Certainly, for $i_b(\tau)<1$, the motion of the phase particle is confined within the potential well, corresponding to the zero-voltage state across the JTD. With the linear increase of the biased current, i.e., $i_b(\tau)\rightarrow 1$, the potential barrier gradually vanishes, yielding the phase particle escapes from the trapped well. As a consequence, a non-zero-voltage signal across the JTD can be detected. 

\begin{figure}[htbp]
\includegraphics[width=0.8\linewidth]{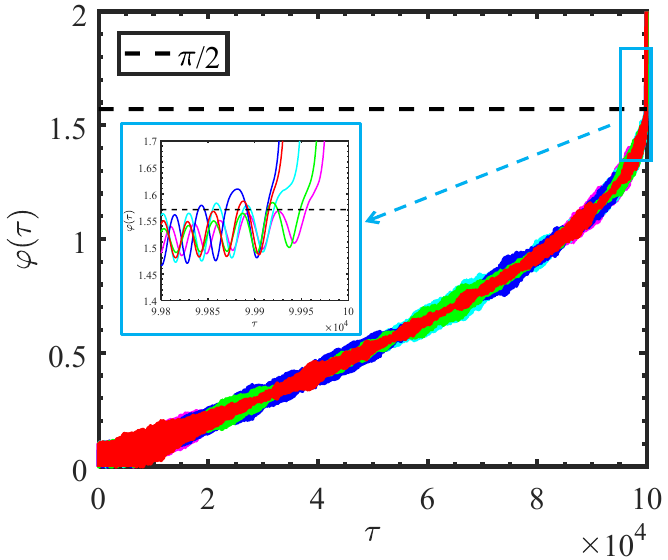}
\caption{The phase-time curves obtained by numerically solving Eq.~\eqref{eq:RCSJ1} five times. Here, the relevant parameters are set as $\varphi_0=0.1$, $\dot{\varphi_0}=0$, $\beta=10^{-4}$,  $v=1\times10^{-5}$ and $2\beta k_BT/E_{J0}=10^{-7}$ (typical thermal noise intensity for a microampere-level JJ at 50 mK with a GHz plasma frequency), respectively. }\label{FF2}
\end{figure}

However, due to noise effects (which can be interpreted as introducing a random driving current to the JTD, thereby reducing the potential barrier height that confines the phase particle), the JTD possesses a finite probability of being switched from the zero-voltage state to a non-zero one, even when $i_b(\tau)<1$~\cite{noise1,noise2} still holds. To verify these events, we numerically solved the driven phase dynamics equation (Eq.~\eqref{eq:RCSJ1}) for the noisy JTD device five times. The results are shown in Fig.~\ref{FF2}, which demonstrates that the noise significantly influences the switching current distribution of the JTD. For the numerical calculations, we adopted a normalized time step size of $\Delta \tau=0.02$ and defined the voltage state critical condition as $\varphi(\tau)>\varphi_{esc}=\pi/2$~\cite{SCD0}. When the JTD phase exceeds this critical value, we assume that the JTD is switched from its zero-voltage state to a non-zero-voltage state, producing a measurable voltage signal across the JTD.
Consequently, a statistic switching current distribution (SCD) could be obtained by the experimentally repeated measurements of the switching currents, which can be fitted by the usual semi-empirical formula~\cite{SCD_eq}  
\begin{equation}
\begin{aligned}
P(i_{sw})=\frac{\Gamma(i_{sw})}{v}\int_0^{i_{sw}}v\Gamma(i'_{sw})di'_{sw}\,,
\end{aligned} \label{eq:SCD}
\end{equation}
where
\begin{equation}
\begin{aligned}
\Gamma(i_{sw})_{th}=\frac{\omega_p}{2\pi}a_{th}\exp\left(-\frac{\Delta U(\varphi)}{k_BT}\right)\,,
\label{eq:TA}
\end{aligned}
\end{equation}
is the escape probability of the phase particle escaping from the potential well, due to thermal activation~\cite{noise2}, and
\begin{equation}
\begin{aligned}
\Gamma(i_{sw})_q=\frac{\omega_p}{2\pi}a_q\exp\left[-\frac{\Delta U(\varphi)}{k_BT}(1+\frac{0.87}{Q})\right]\,.
\label{eq:Q}
\end{aligned}
\end{equation}
refers to that due to the quantum tunneling~\cite{Leggett}. The parameters in Eqs.~(\ref{eq:TA}-\ref{eq:Q}) are usually defined as: $0<a_t\leq1$, $a_q=\sqrt{864\pi\Delta U(\varphi)/\hbar\omega_p}$, $\omega_p=\omega_J(1-i_b^2)^{1/4}$, and $Q=1/\omega_pRC$, respectively. 

Given that the dynamics simulated above are insensitive to the initial phase of the JTD driven by a slowly variable biased current and its noise-dependent observable effects can be effectively fitted by the semi-empirical formulas~\cite{detector4,JJph4,SW1,SW2,SW4}, it had been widely applied to construct various JTDs for detecting the weak microwave signals~\cite{JJph1,JJph4,JJPh8}. In particular, by optimizing the physical parameters of the JTD, such as the capacitance, resistance, and its Josephson energy, we demonstrated in Ref.~\cite{our1} that the JTD with the slowly increasing biased current can be really utilized to implement the weak microwave signals detection, and the achievable detection sensitivity can arrive at a few photons level, approaching the energy quantum limit of the microwave signal. The feasibility of this approach was experimentally verified in Ref.~\cite{our2}. The demonstrated detection sensitivity was really limited by the noise background.
 
A natural question is, how slow is the bias current sweep rate to ensure that the above phase dynamics are independent of the initial phase, hereby the above semi-empirical formula, i.e., the Eq.~\eqref{eq:RCSJ1}, can well fit the experimentally measured SCD? 
Mathematically, Eq.~\eqref{eq:RCSJ1} is a typical nonlinear evolution equation, which implies that the dynamic evolution behavior of the JTD should depend on the initial phase $\varphi_0=\varphi(\tau=0)$)~\cite{initial,SCD0,nonequilibrium1}, besides the physical parameter $\beta$. Below, we answer this question by numerically solving Eq.~\eqref{eq:RCSJ1} for the JTD being biased by the variable current with different sweep rates.

\subsection{Phase dynamics for the JTD biased by the fast variable current: initial phase sensitivity}

In this subsection, we investigate the phase dynamics for the JTD biased by the current with different sweep rates. It is seen clearly from Eq.~\eqref{eq:RCSJ1} that, apart from the noise current $i_n$, the key parameters influencing the phase dynamical behavior are: the dissipation coefficient $\beta$, the sweep rate $v$ of the bias current, and the initial phase $\varphi_0$-of the JTD that has been received relatively little attention in previous studies~\cite{our1,our2,detector7,JJph6,JJph7}. By tuning the sweep rate $v$ of the bias current, one can control the phase dynamical evolution path of the device. Meanwhile, the initial phase $\varphi_0$ defines the starting point of this evolution. Crucially, different initial phase values may lead to distinct evolutionary trajectories.

\begin{figure}[htbp]
\includegraphics[width=0.8\linewidth]{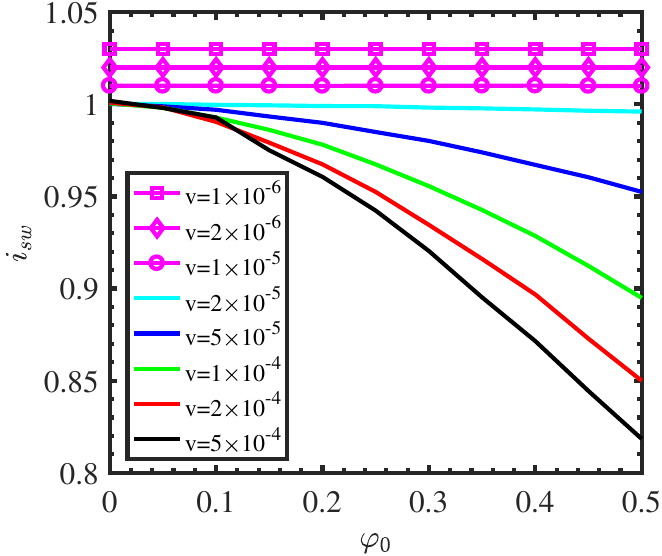}
\caption{The initial phase dependence of the switch currents for the JTD driven by the biased currents with different sweep rates, in the absence of thermal noise. It is noted that the three orange lines are numerically identical. For clearer visualization, they have been vertically offset by 0.03, 0.02, and 0.01 (from top to bottom), respectively. The other parameters are set the same as in Fig.~\ref{FF2}.}\label{FF3}
\end{figure}

To simply demonstrate the initial phase sensitivity of phase evolution dynamics, we first numerically solve the noiseless dynamical equation~\eqref{eq:RCSJ1} with $i_n(\tau)=0$ for different sweep rates $v$ and initial phase $\varphi_0$. For convenience, we assume that the energy dissipation coefficient $\beta$ is constant. The numerical results of the switch currents of the JTD with the above selected parameters are shown in Fig.~\ref{FF3}.
One can see clearly that, if the dimensionless sweep rate of the biased current is sufficiently small (e.g., $v<1\times 10^{-5}$), the switching currents of the JTD are independent of the values of $\varphi_0$. This switching behavior can be well fitted by the conventional semi-empirical formula~\eqref{eq:SCD}. Interestingly, as the sweep rate increases, different initial phase values lead to different switching currents. This is the initial phase sensitivity of the phase dynamics. Obviously, the data related to $\varphi_0=0$ should be deleted, as it corresponds to the vanished Josephson effect. Importantly, it is seen that the switching currents of the JTD can be strongly influenced by the non-zero initial phase for a given device with the deterministic $\beta$-parameter. 

To quantify the relation between the dissipation coefficient of the JTD and the sweep rate of its biased current, we introduce an effective sweep rate parameter: $\kappa=v/\beta$. In Tab.~\ref{tab:table1} we list its values, corresponding to the numerical results shown in Fig.~\ref{FF3}, for the JTD with $\beta=1\times 10^{-4}$. 
\begin{table}[htbp]
\caption{\label{tab:table1}The effective bias current sweep rate parameter $\kappa=v/\beta$ is defined by the dissipation coefficient $\beta$ and the bias current sweep rate $v$ in Fig.~\ref{FF3}, for $\beta=1\times 10^{-4}$.}
\begin{ruledtabular}
\begin{tabular}{cccc}
$v$& $\kappa=v/\beta$& $v$& $\kappa=v/\beta$\\
\hline
 $1\times10^{-6}$&0.01 & $5\times10^{-5}$&0.5 \\
 $2\times10^{-6}$&0.02 & $1\times10^{-4}$&1 \\
 $1\times10^{-5}$&0.1 & $2\times10^{-4}$&2 \\
 $2\times10^{-5}$&0.2 & $5\times10^{-4}$&5\\
\end{tabular}
\end{ruledtabular}
\end{table}
Depending on the magnitude of this effective sweep rate parameter, JTD can be treated as being ``adiabatic" or ``non-adiabatic" driven, see Appendix~\ref{sec:appendix.A} for details. Correspondingly, JTD can be considered to be in two working states: One is in equilibrium, and the other is in a non-equilibrium state. As a consequence, the phase dynamics of the JTD working in different working states exhibit different characteristics.

i) The phase dynamics for the equilibrium JTD are the initial phase independence. 

If the effective current sweep rate is sufficiently slow, i.e., $\kappa\ll 1$, the JTD works at its equilibrium state~\cite{nonequilibrium1,nonequilibrium2}. It is seen from Fig.~\ref{FF3} that the switch current of such a JTD is always equivalent to 1, i.e., the critical one of the JTD, whatever the initial phase of the JTD is. This indicates that the sufficiently low sweep rate enables the phase particle to rapidly stabilize near the local minimum $\varphi_{min}$ of its trapped potential well, as the adiabatically increased vibrational energy of the phase particle can be quickly offset by its dissipated energy, keeping the JTD works in its equilibrium state. Therefore, the phase dynamics of the equilibrium JTD must be independent of the value of its initial phase. In this case, the value of the $\beta$-parameter plays an important role, which implies that the thermal noise must influence the achievable detection sensitivity of the equilibrium JTD applied for weak signal detection.     

ii) The phase dynamics for the non-equilibrium JTD is the initial phase sensitivity.

Conversely, if the effective current sweep rate satisfies the condition $\kappa\geq 1$, i.e., the JTD is driven non-adiabatically and works at its non-equilibrium state~\cite{nonequilibrium1,nonequilibrium2}, Fig.~\ref{FF3} shows that its switching current becomes dependent on the initial phase of the JTD. This is because that, under the fast driving the phase particle can not be stabilized near the previous $\varphi_{min}$ point, as the phase particle can not stabilize near the previous $\varphi_{min}$ point and thus the energy equilibrium between the increasing driven energy and the lost dissipated energy has been broken, yielding the stronger nonlinearity of the phase dynamics. Macroscopically, as shown in Fig.~\ref{FF3}, the observable switch current $i_{sw}$ lowers with the increased sweep rate $v$ and the initial phase $\varphi_0$. 

Physically, the dissipation factor $\beta$ could originate from certain noise resources of the vibrating phase particle, and a fast applied biased current could be treated as the excited resource of the vibrational phase particle in the trapped well~\eqref{eq:potential}. Therefore, with the increase of the sweep rate $v$ of the biased current, the dynamics of the phase particle can be immune to the noises and sensitive to the value of its initial phase. 
Given that the noise influence has been effectively suppressed, the non-equilibrium JTD should possess a higher detection sensitivity over the equilibrium JTD for the weak signal detection. This argument would be numerically verified in the next section.

\subsection{Thermal noise characteristics of SCDs for the JTD working in either equilibrium or non-equilibrium state}
With the experimentally measurable SCD of a JTD, we now numerically simulate its observable effect related to the thermal noise and initial phase, when it works at different states, i.e., either the equilibrium or non-equilibrium state. Still, we only consider the thermal noise for simplicity.

\begin{figure}[htbp]
\includegraphics[width=1\linewidth]{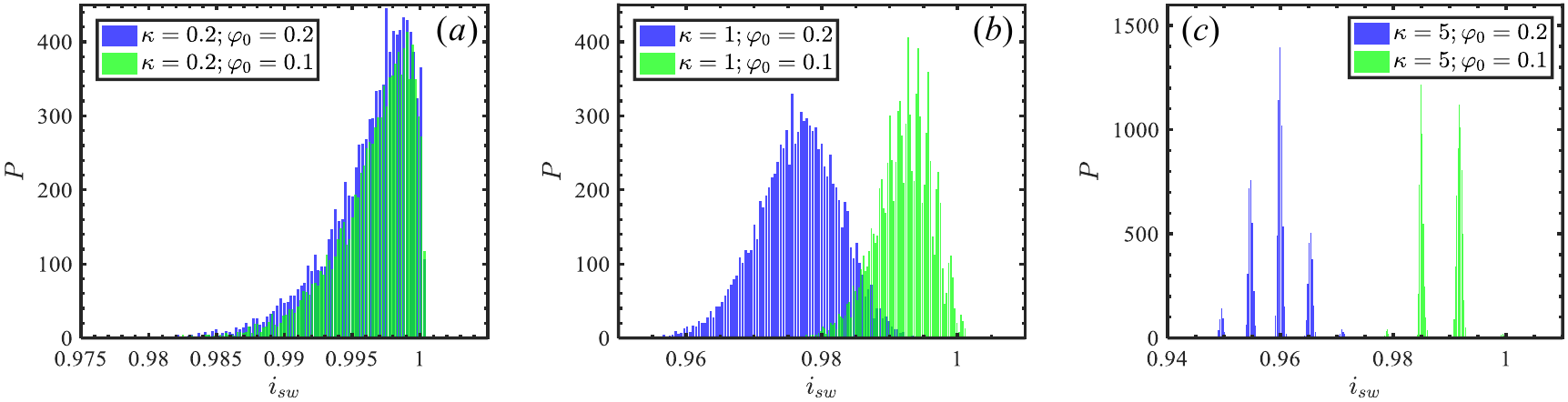}
\caption{Numerical simulations of the SCDs for a JTD driven by the biased currents with the typical sweep rates. The other parameters are set the same as in Fig.~\ref{FF2}.}\label{FF4}
\end{figure}

\begin{figure}[htbp]
\includegraphics[width=0.8\linewidth]{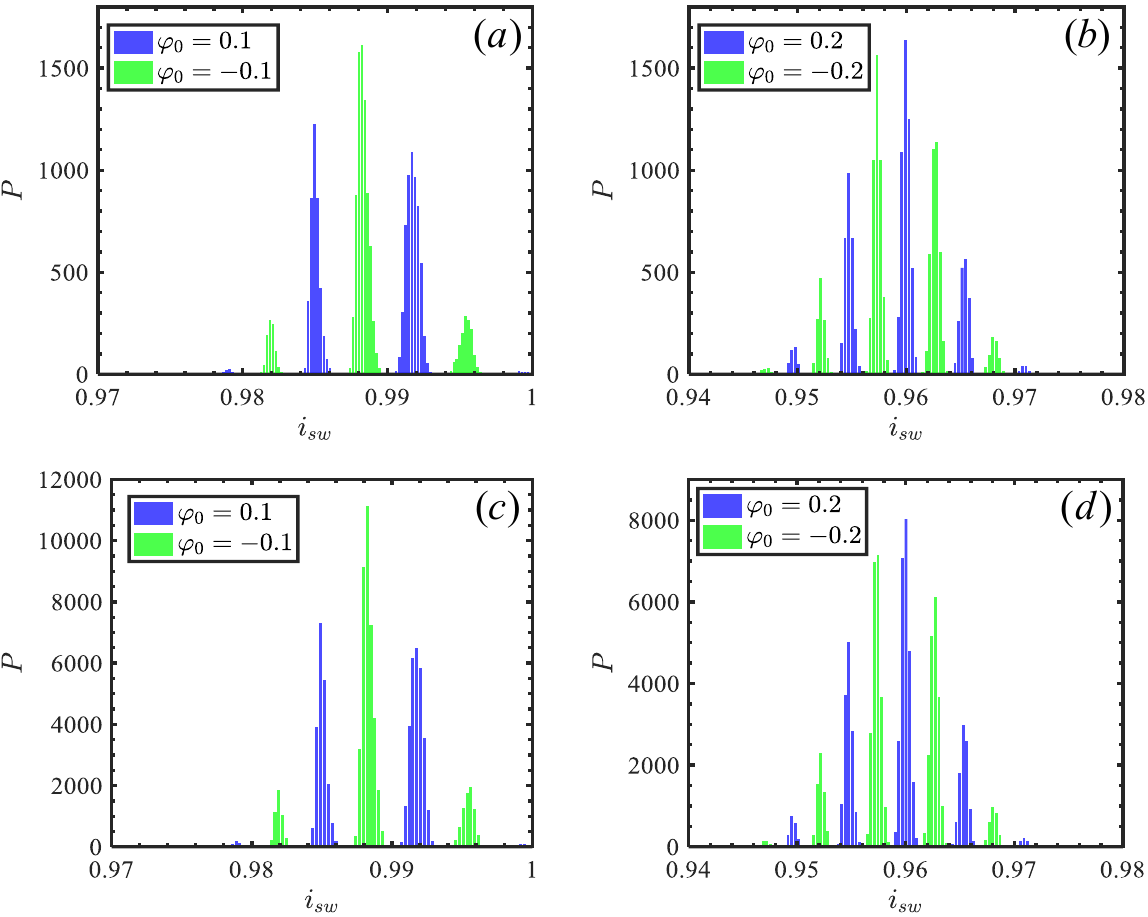}
\caption{The numerical simulation results of JTD's SCD for different initial phases. The parameters are set as $\kappa=5$ and the other parameters are set the same as in Fig.~\ref{FF2}. The initial phase is set to $\varphi_0=\pm0.1$ in (a) and (c), and to $\varphi_0=\pm 0.2$ in (b) and (d). In the figure, (a) and (b) show the statistical distribution from 10,000 simulation runs, while (c) and (d) correspond to 50000 simulation runs.}\label{FF5}
\end{figure}

For the comparison, we first numerically demonstrated that the SCD of an equilibrium-state JTD, i.e., it is biased by a sufficiently slow variable current, is sensitive not to the initial phase. Specifically, for each of the initial phase $\varphi_0=0.1$ and $\varphi_0=0.2$, we numerically solved Eq.~\eqref{eq:RCSJ1} 10000 times for the given sweep rate of $\kappa=0.2$ and the thermal noise with $2\beta k_BT/E_{J0}=10^{-7}$ and then recorded the relevant SCDs in Fig.~\ref{FF4} (a). Notably, the simulated SCD shown in Fig.~\ref{FF4}(a) agrees well with the experimentally observed SCD data~\cite{SW1,RCSJ3}, i.e., the SCD of an equilibrium-state JTD is insensitive to its initial phase.
However, as the numerical results shown in Fig.~\ref{FF4}(b, c), when the sweep rate increases, typically, e.g., $\kappa=1$, and $5$, the SCDs become obviously initial phase dependent. This indicates that the observable SCDs of the non-equilibrium JTD show the pronounced initial phase dependence, even when the thermal noise exists. With the increase of sweep rate, the more manifest splitting feature of the SCD can be observed, yielding the stronger initial phase dependence. To further verify the above observations, with the same sweep rate, i.e., $\kappa=5$, the SCDs for $\varphi_0=\pm 0.1$ are also shown in Fig.~\ref{FF5}, wherein (a-b) and (c-d) represent the results obtained by repeating 10,000 and 50,000 times, respectively. One can see that the number of split peaks in the SCD is only positively correlated with the absolute value of the initial phase $\varphi_0$. And, the SCDs do not overlap for the two initial phases that are opposite in sign, i.e., for $\varphi_0$ and $-\varphi_0$.

The above numerical results indicate that a fast variable biased current can ``non-adiabatically" drive the JTD into its non-equilibrium state, wherein its phase dynamics are sensitive to the initial phase, and the thermal noise influence could be effectively suppressed. In appendix~\ref{sec:appendix.B}, such a thermal noise insensitivity was further verified by the relevant numerical experiments. Fortunately, the initial phase of the JTD is feasible. In appendix~\ref{sec:appendix.C}, we demonstrated a conventional approach to experimentally implement the accurate modulation of the initial phase $\varphi_0$. Interestingly, the initial phase sensitivity of phase dynamics can serve as a distinctive ``fingerprint" in the measured SCD for the non-equilibrium JTD. Since this fingerprint in SCD is not affected by noise, a method based on identifying small changes in fingerprints can be constructed to achieve the expected ultra-high sensitivity weak signal detection. In what follows, we will demonstrate its feasibility and numerically estimate its performance indicators.  

\section{\label{sec:3}Implementation of the microwave single-photon detection by using a non-adiabatically driven JTD}
Based on the above analysis of the phase dynamics for a JTD with the adjustable current biases and its observable effects related to the initial phase and thermal noise, we now demonstrate that a non-equilibrium JTD, wherein thermal noise can be effectively suppressed, can be experimentally implemented for the highly sensitive microwave weak signal detection. Its achievable detection sensitivity can arrive at the single microwave-photon level.
\subsection{Detection Method}
The vast majority of previous studies for microwave weak signal detection by using the JTD devices, including our recent work~\cite{our1} and those published in Refs.~\cite{detector4,JJph2,JJph4,JJph6}, had primarily focused on how to improve the achievable detection sensitivity by reducing the noise of the JTD device.
While noise can never be completely eliminated for any realistic detection device, its achievable detection sensitivity is always limited by the background noise~\cite{our1}. In fact, in detection schemes, the bias current of the JTD is applied sufficiently low (i.e., adiabatically) for maintaining the detector working in its equilibrium state, thereby the influence from the noise is always-on~\cite{detector4,detector7,JJph1,JJph2,JJph4,JJph5,pankratov_detection_2025}. Differently, as we demonstrated above, the rapidly changing bias current can non-adiabatically drive the JTD into the non-equilibrium state, wherein its phase dynamical evolution process could be insensitive to thermal noise. This suggests that non-equilibrium JTD might possess higher sensitivity, over its equilibrium-state counterpart, for the microwave weak signal.

\begin{figure}[htbp]
\includegraphics[width=1\linewidth]{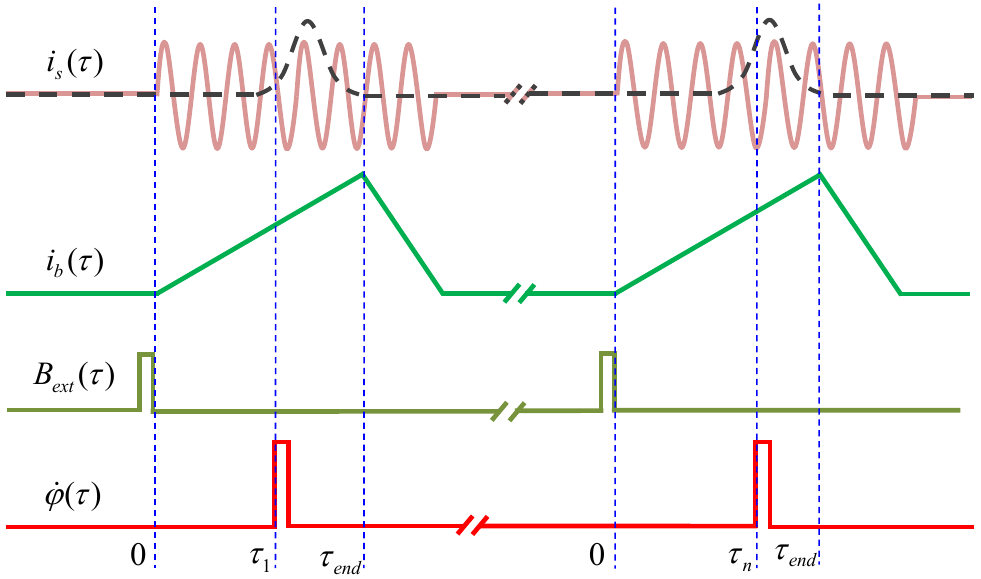}
\caption{Schematic diagram of a microwave signal detection scheme using a non-equilibrium JTD. The vertical axis represents the operational time sequences. From top to bottom, the panels illustrate the input microwave signal, the triangular bias current applied to the JTD, the external magnetic field applied to modulate the initial phase of the JTD, and the real-time monitoring of the non-zero voltage signals across the JTD. }\label{F7}
\end{figure}

Physically, the process for an equilibrium JTD being used to implement the weak microwave signal $i_s(\tau)$ can be described by the following equation~\cite{detector4,JJph2,JJph4,JJph6,JJph7,JJPh8,our1,our2}
\begin{equation}
\frac{{\rm d}^2\varphi}{{\rm d}\tau^2}+\beta\frac{{\rm d}\varphi}{{\rm d}\tau}+\sin(\varphi)=v\tau+i_n(\tau)+i_s(\tau)\,,
\label{eq:detection}
\end{equation}
where the bias current $i_b(\tau)$ is specifically set as a rapidly increasing linear current, i.e., $i_b(\tau)=v\tau$ with the sweep rate $v$ being significantly high, i.e., $v/\beta\geq 1$. As a consequence, the JTD is driven into its non-equilibrium state to ensure the corresponding phase dynamics are immune to the noise but highly sensitive to the initial phase. An experimental configuration by using a non-equilibrium JTD to implement the single microwave-photon detection could be designed in Fig.~\ref{F7}, wherein the continuous or pulsed microwave signals to be detected are represented by the pale red curve and the black curve, respectively. The bias current of the JTD is marked as the green curve; a rapidly swept periodic triangular current signal is applied to drive the JTD into its non-equilibrium state. The magnitude of such a signal increases linearly from zero with each cycle until a detectable voltage switch occurs in the JTD. The magnetic field signal $B_{ext}(\tau)$, indicated by the olive color and synchronized with the bias current, is applied to modulate the initial phase of the JTD. In order to avoid the potential voltage state transitions in the JTD, its value should be much smaller than the critical magnetic field of the JTD. Within each bias current cycle, the time zero is defined as the initial moment when the triangular bias current is applied. Thus, the switching time of the JTD voltage state is marked as the moment when the voltage signal appears across the JTD. 

With the method shown in Fig.~\ref{F7}, the progress for the microwave signal detection can be performed sequentially as:

i) measure first the SCD of the JTD without the signal inputs, i.e., $i_s(\tau)=0$, and mark it as $P_0=(SCD)_0$;

ii) measure another the SCD of the JTD with the signal inputs, i.e., $i_s(\tau)\neq 0$, mark it as $P_1=(SCD)_1$;

iii) provide the judgment of whether there is a microwave signal input on the data processing side, by realizing the distinction between $P_1$ and $P_0$.  

Apparently, the present method to implement the weak microwave signal detection differs from the conventional one implemented by using the equilibrium JTD in two key aspects: First, the bias current sweep rate is significantly higher to drive the JTD into a non-equilibrium state, called the non-equilibrium JTD, wherein the noise effect has been effectively isolated. Second, an additional offline adjustable DC current is applied to generate a tunable external magnetic field for the flexible setting of the initial phase, i.e., to get the SCD fingerprint (i.e., $P_0$) of the non-equilibrium JTD. Subsequently, the input signal can be detected by identifying whether such a fingerprint was changed.

Importantly, differing from the usual single-peak SCDs obtained by the equilibrium JTD~\cite{JJph2,JJph4} (and thus their identification can be implemented by calculating the $d_{KC}$-index~\cite{JJPh8,our1,our2,KC1}), the SCD fingerprints obtained by the present non-equilibrium JTD, i.e., both the background $P_0$ and the signal $P_1$, are multi-peaks. Indeed, distinguishing between the two statistical distributions constitutes a binary classification problem, which aims to determine whether $P_0$ and $P_1$ belong to different categories~\cite{ROC1,ROC2,ROC3,classify}.
\begin{figure}[htbp]
\includegraphics[width=0.9\linewidth]{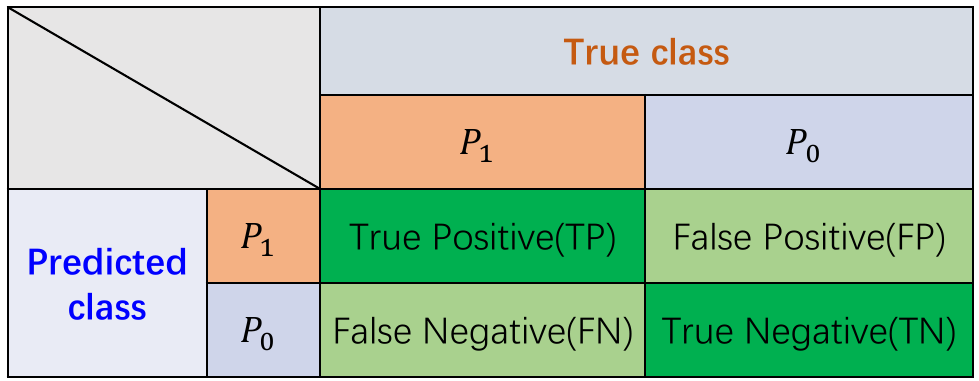}
\caption{The confusion matrix formed by the SCDs; $P_0$ and $P_1$ in the absence and presence of the incident microwave signal, respectively.}\label{F8}
\end{figure} 
According to the binary classification method in statistics, four distinct outcomes for implementing the discrimination between $P_0$ and $P_1$ can be simply illustrated in Fig.~\ref{F8}, wherein 

a) the true positive (TP)  subset. It represents that the probability for the SCD being predicted in the $P_1$, and it indeed belongs to the $P_1$. Thus, the prediction is correct.

b) the false positive (FP) subset, which implies that the probability of the SCD being predicted as the $P_1$ events, but they actually belong to the $P_0$. Thus, the prediction is incorrect.

c) the false negative (FN) subset. It refers to the probability of the SCD being predicted to be in $P_0$, but it actually belongs to the $P_1$. Then, the prediction is incorrect;

d) the false positive (FP) subset, which represents the probability of the SCD being predicted in the $P_0$, and it indeed is in the $P_0$. Then, the prediction is correct.

\begin{figure}[htbp]
\includegraphics[width=0.8\linewidth]{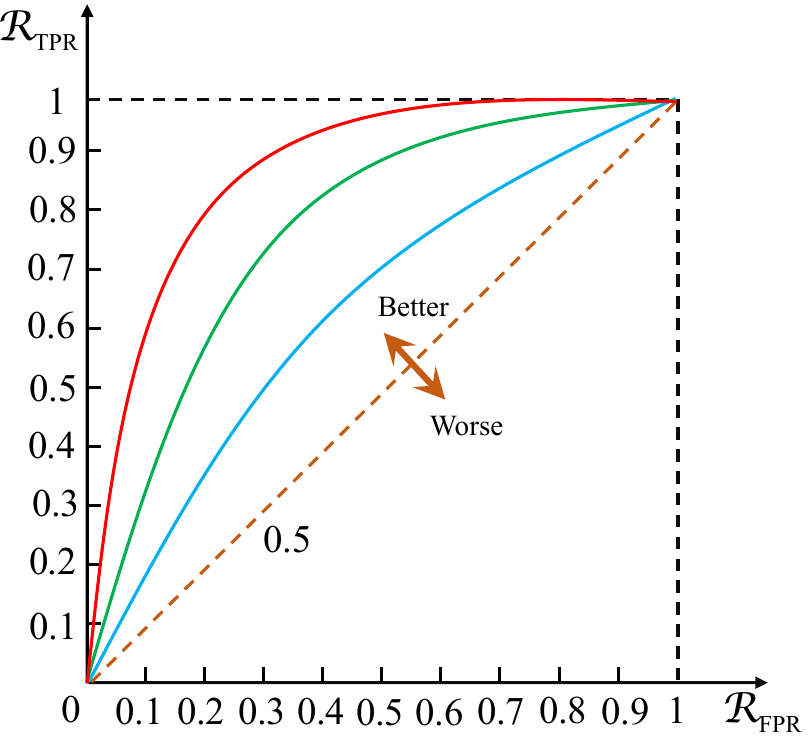}
\caption{A schematic diagram of the ROC curve for the $P_0$-SCD and $P_1$-SCD. }\label{F9}
\end{figure}

Statistically, the accuracy of such a binary classification can be quantified~\cite{ROC1,ROC2,ROC3} by using a ROC curve and calculating its AUC value, $\mathcal{R}_{\rm AUC}$, to distinguish the $P_0$ and $P_1$. As shown in Fig.~\ref{F9}, the ROC curve is a graphical tool to characterize the distinguishability between $P_0$ and $P_1$. The vertical and horizontal axes respectively, refer to the probabilities of the True Positive Rate (TPR) and False Positive Rate (FPR). The TPR is defined as the proportion of instances that truly belong to $P_1$ and thus its value can be calculated as~\cite{ROC2,ROC3,classify}; $\rm \mathcal{R}_{TPR}=TP/(TP+FN)\,$. Similarly, the FPR is defined as the proportion of instances that truly belong to $P_0$ but are incorrectly predicted as belonging to $P_1$. Its value can be calculated as~\cite{ROC1,ROC2,ROC3} $\rm \mathcal{R}_{FPR}=FP/(FP+TN)\,$.
Clearly, either the higher value of $\mathcal{R}_{\rm TPR}$ or the lower value of $\mathcal{R}_{\rm FPR}$ corresponds to the better distinguishability between $P_0$ and $P_1$. 
A parameter known as the AUC (Area Under the Curve)~\cite{ROC2,ROC3,classify};
\begin{equation}
\begin{aligned}
\mathcal{R}_{\rm AUC}=\int_{0}^{1}\mathcal{R}_{\rm TPR}(\mathcal{R}_{\rm FPR})d(\mathcal{R}_{\rm FPR})\,.
\end{aligned}
\end{equation}
which refers to the area under the ROC curve, is commonly introduced to quantify the distinguishability between two statistical distributions. The value of $\mathcal{R}_{\rm AUC}$ is closer to 1, indicating the stronger distinguishability between the two distributions. Certainly, $\mathcal{R}_{\rm AUC}=0.5$ suggests the completely indistinguishability. Specifically, in Fig.~\ref{F9}, the values of $\mathcal{R}_{\rm AUC}$, corresponding to the red, green, and blue ROC curves correspond progressively decrease, and finally the orange dashed line represents the complete indistinguishability with $\mathcal{R}_{\rm AUC}=0.5$. Of course, the $\mathcal{R}_{\rm AUC}$ of any distribution compared with itself is exactly 0.5, since any distribution is inherently indistinguishable from itself. 

Experimentally, we can combine n measured switching currents from both $P_0$ and $P_1$ into a single array:
\begin{equation}
\begin{aligned}
X=[i_1,i_2,\dots,i_{2n}], i_1\leq i_2\leq\dots\leq i_{2n}\,,
\end{aligned}
\end{equation}
which is then sorted in ascending order.
The ROC curve and the value of $\mathcal{R}_{\rm AUC}$ can be obtained as follows: 

i) treat each value of in X as a threshold $\theta$ and calculate the proportion of switching currents in $P_1$; if they are greater than or equal to $\theta$, we get the $\mathcal{R}_{\rm FPR}$. Similarly, compute the proportion of values in the $P_0$, we get the $\mathcal{R}_{\rm FPR}$ for those exceeding $\theta$. 

ii) for each threshold $\theta$, we calculate the $(\mathcal{R}_{\rm FPR}$ and $\mathcal{R}_{\rm TPR})$ and then provide its ROC curve. 

iii) calculate the value of $\mathcal{R}_{\rm AUC}$ to quantitatively characterize the distinguishability between the distributions $P_0$ and $P_1$, thereby enabling the evaluation of the non-equilibrium JTD  detectability for weak signal detection. 

As the calculation of the $\mathcal{R}_{\rm AUC}$-parameter does not require the assumption of regarding the underlying data distribution or sample size, this method is particularly suitable for distinguishing various multi-peak distributions, specifically here for the $P_0$ and $P_1$ obtained by the non-equilibrium JTD for weak microwave signals detection.
\subsection{\label{sec:3B}Numerical Simulations}

To validate the effectiveness of the method proposed above, below we perform a series of numerical experiments by solving Equation~\eqref{eq:detection} 10,000 times for the non-equilibrium JTD with and without a significantly weak microwave signal input, thereby obtaining the $P_1$ and $P_0$ SCDs, respectively. Then, by quantifying the distinguishability between $P_1$ and $P_0$, we demonstrate that the applied non-equilibrium JTD can be utilized to implement the desired  single microwave-photon detection, i.e., its achievable detection sensitivity can really arrive at the single-photon level in the microwave band.

\begin{figure}[htbp]
\includegraphics[width=1\linewidth]{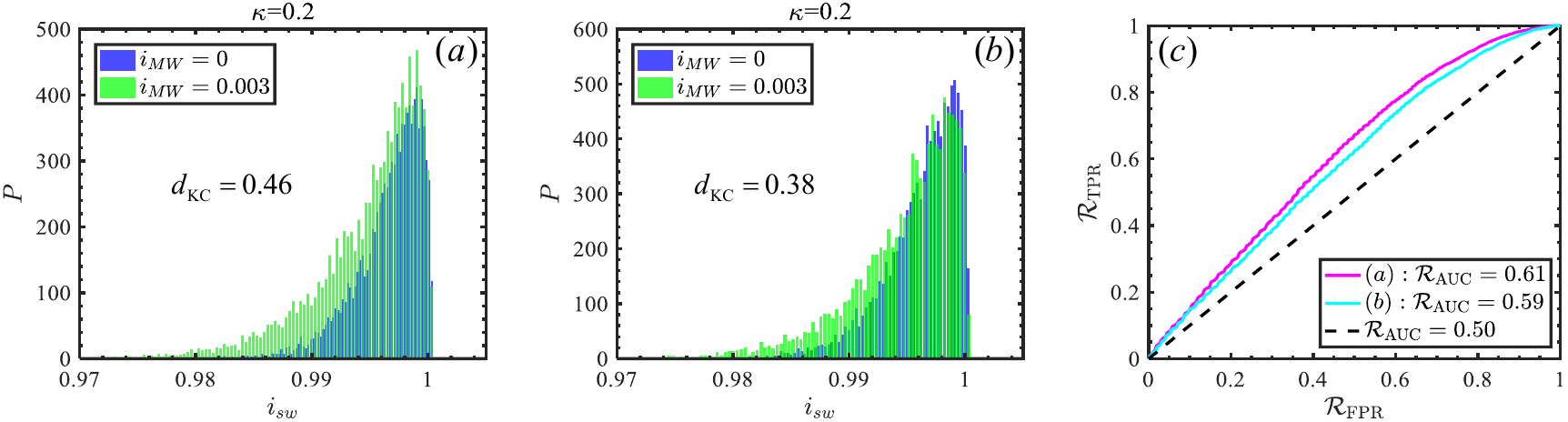}
\caption{The simulated distinguishability of the SCDs for the equilibrium JTD, with the different initial phases. The $d_{KC}$-indexes~\cite{our1} shown in subfigures (a) and (b) indicate that the input signals are detectable, while the values of the $\mathcal{R}_{\rm AUC}$-parameter shown in subfigures (c) indicate that their detectabilities are significantly low. Here, the relevant parameters are set as: $\omega=1$, (a,b) for the initial phase is $\varphi_0=0.1$ and $\varphi_0=0.2$, and the other parameters are set the same as in Fig.~\ref{FF2}.}\label{F10_1}
\end{figure}

\begin{figure}[htbp]
\includegraphics[width=1\linewidth]{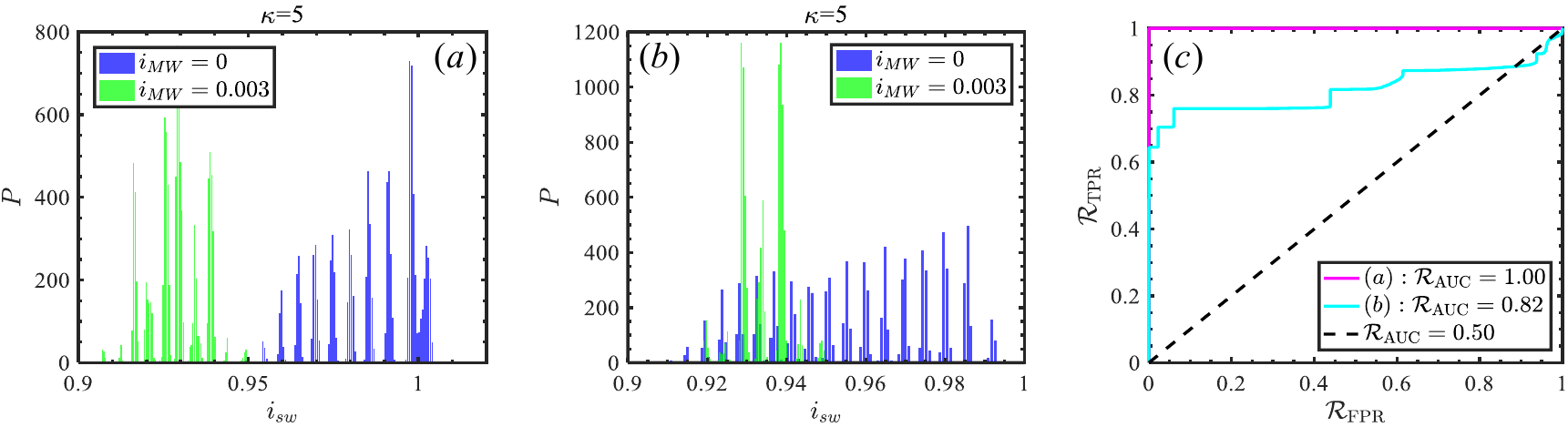}
\caption{Numerical simulations of the signal delectabilities by using the non-equilibrium JTD with different initial phases. Subfigures (a) and (b) show the changed SCDs with and without the signal input, while subfigure (c) shows their distinguishabilities. }\label{F10_2}
\end{figure}

First, we visually demonstrate the sensitive advantage of the non-equilibrium JTD over its equilibrium counterpart for the weak microwave signal detection. Without loss of generality, we consider a continuous-wave weak microwave signal, which is treated as a current signal:
\begin{equation}
\begin{aligned}
i_s(\tau)=i_{MW}\sin(\omega_{MW} \tau)\,,\label{eq:MWcurrent}
\end{aligned}
\end{equation}
through the JTD, where $i_{MW}$ and $\omega_{MW}$ denote the amplitude and frequency of the signal, normalized to the JTD critical current $I_c$ and the plasma frequency $\omega_J$, respectively. 
Specifically, by numerically solving the relevant phase dynamics equation~\eqref{eq:detection}for the equilibrium JTD with the bias current sweep rate $\kappa=0.2$, Figs.~\ref{F10_1}(a-b) present the corresponding SCDs with and without the microwave signal input~\eqref{eq:MWcurrent}. Following Ref.~\cite{our1}, the calculated $d_{KC}$-indexes for the given thermal noise show that such a relatively strong microwave signal is detectable by using the equilibrium JTD. However, as shown in the Figs.~\ref{F10_1}(c) for their values of $\mathcal{R}_{\rm AUC}$ of the corresponding ROC curves, the distinguishability between their SCDs is really low and also insensitive to the initial phases, due to the influence of the thermal noise.
The situation would be very different if the JTD were working at its non-equilibrium state implemented by using the significantly high sweep rate, i.e., $\kappa\geq 1$. Specifically, by numerically solving the phase dynamics for the non-equilibrium JTD with $\kappa=5$ in the same thermal noise environment, we get the relevant SCDs Figs.~\ref{F10_2}(a, b), for the initial phase being set as $\varphi_0=0.1$ and $\varphi_0=0.1$, respectively. One can see that, for a given initial phase of a non-equilibrium JTD, the input signal modifies its SCD fingerprint and thus its detectability can be quantified by its ROC curves and the $\mathcal{R}_{\rm AUC}$ values. For a given input signal, we have $\mathcal{R}_{\rm AUC}=0.82$ and $1.00$ respectively for the ROC curves with $\varphi_0=0.1$ and $\varphi_0=0.2$. This clearly demonstrated that the non-equilibrium JTD possesses a stronger ability for weak signal detection, as its phase dynamics have achieved the effective isolation of noise.  
As a consequence, the SCDs with and without the signal input possess higher distinguishability, compared to those in their equilibrium counterpart. Therefore, given the equilibrium JTD can provide the detection of microwave signal with the sensitivity approaching the single-microwave-photon energy level~\cite{our1}, using the non-equilibrium JTD to implement the more sensitive detection of microwave signal up to its single-photon level, should be feasible.

\begin{figure}[htbp]
\includegraphics[width=1\linewidth]{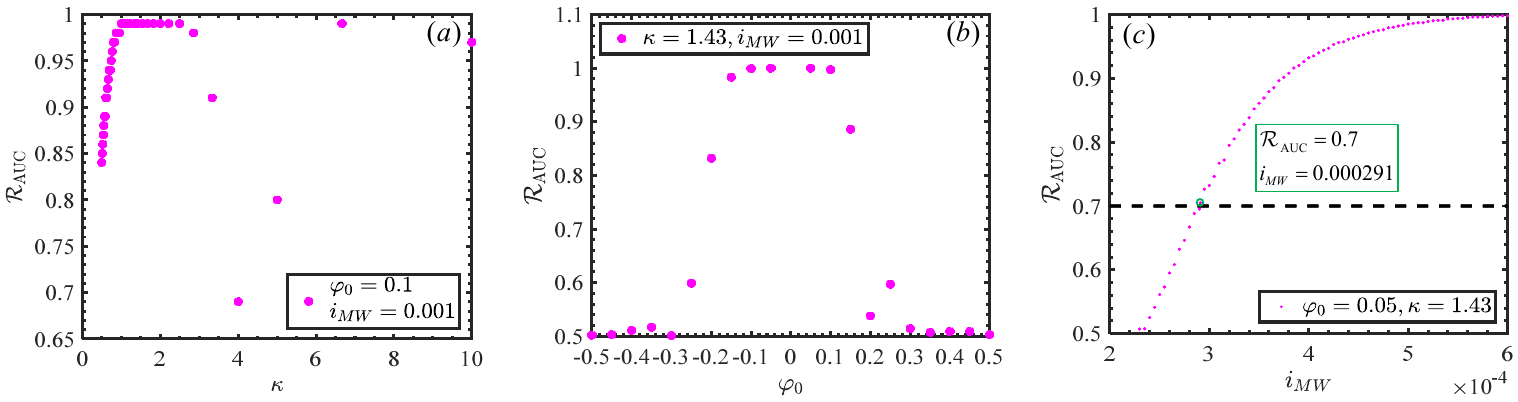}
\caption{Numerically optimized the operational parameters of a
non-equilibrium JTD to ensure it possesses the detection
ability of the signal with a single photon energy; (a) for the $\kappa$-parameters, (b) for the $\varphi_0$-parameter, and (c) for the signal intensity, respectively. Here, we set $\omega_{MW}=1$ and the other parameters are set the same as in Fig.~\ref{FF2}.}\label{F11}
\end{figure}

Next, we specifically demonstrate how to optimally set the sweep rate $\kappa$ and the relevant initial phase $\varphi_0$ of a non-equilibrium JTD to ensure it possesses the detection ability of the signal with a single photon energy. Typically, for the input of a continuous-wave weak microwave signal with $i_s=0.001$ and $\omega_{MW}=1$, Fig.~\ref{F11}(a) shows how the value of $\mathcal{R}_{\rm AUC}$-index depends on the $\kappa$-parameter of a non-equilibrium JTD, whose initial phase is beforehand set as $\varphi_0=0.1$. Simultaneously, for such a signal input, Fig.~\ref{F11}(b) demonstrates how the value of $\mathcal{R}_{\rm AUC}$-index depends on the initial phase $\varphi_0$-parameter for a non-equilibrium JTD with the optimized $\kappa$-parameter  $\kappa=1.43$. Based on these numerical optimizations, we designed a non-equilibrium JTD with $\kappa=1.43$ and $\varphi_0=0.05$, and showing in Fig.~\ref{F11}(c) its achievable value of the $\mathcal{R}_{\rm AUC}$-index increasing with the intensity of the input microwave signal.
It can be observed that the achievable $\mathcal{R}_{\rm AUC}$, representing the reliable distinguishability of the $P_1$-SCD from the $P_0$-SCD, increases really with the intensity of the incident microwave signal. With the arguments based on the statistical decision rules~\cite{ROC3}, i.e., the input microwave weak signal can be considered to be reliably detected by the designed non-equilibrium JTD, if its achievable value of $\mathcal{R}_{\rm AUC}$ is not less than $0.7$, we show the weakest detectable continuous-wave microwave signal, i.e., $i_{MW}=0.000291$, in Fig.~\ref{F11}(c) by using the designed non-equilibrium JTD with $\beta=10^{-4}$ in the noise environment $2\beta k_BT/E_J=10^{-7}$. Consequently, following Ref.~\cite{PhysRevApplied.22.024015}, the corresponding minimum detectable continuous-wave microwave power is estimated as 
\begin{equation}
\begin{aligned}
P_{min}=\frac{i_{MW}^2I_c^2R_{MW}}{2\chi}\approx8.4681\times10^{-6}I_c^2\,,
\end{aligned}\label{eq:minpower}
\end{equation}
wherein the microwave absorbed efficiency is set as $R_{MW}=100~\Omega$ and the impedance of the used JTD is set as $\chi=0.5$ to satisfy the usual impedance-matched condition, respectively.
AS the critical current $I_c$ of the Josephson junction used for the JTD could be typically lowered to a few nanoampere (nA) level, the corresponding minimum detectable power of the non-equilibrium JTD can be lowered to $10^{-23}$~W order. This implies that the non-equilibrium JTD with the optimized parameters could be applied to implement the microwave single-photon detection, as its achievable detection sensitivity can reach the desired microwave single-photon energy level.   

Thirdly, let us validate the practical feasibility of this potential by numerically simulating the detectability of a pulsed single-photon microwave signal, which had been illustrated in Fig.~\ref{F7}. This is because pulsed microwave currents can be expressed in the form of the number of microwave photons~\cite{pulse}.
\begin{equation}
\begin{aligned}
i_{ ph}(\omega_{ ph}, \tau_{ ph}, \tau)=&\sqrt{N_{ph}}i_{ph}\exp\left(-\frac{1}{2}(\frac{\tau-\tau_{\rm d}}{\tau_{ ph}})^2\right)\\
&\times\cos(\omega_{ ph}(\tau-\tau_{ d}))\,,\label{eq:ph}
\end{aligned}
\end{equation}
where $N_{ph}$ denotes the number of microwave photons contained in the pulse current, $\omega_{ph}$ represents the frequency of the incident microwave photon normalized to the Josephson plasma frequency $\omega_J$, $\tau_{d}$ and $\tau_{ph}$ are the arrival time and width of the pulse, respectively. Noted that in Eq.~\eqref{eq:ph}, $i_{ph}=I_{ph}/I_c=\sqrt{\hbar\omega_{\rm ph}\omega_J^2/RI_c^2\tau_{\rm ph}}$ is the effective amplitude of the microwave pulse current normalized to the critical current $I_c$ of the JTD.

\begin{figure}[htbp]
\includegraphics[width=1\linewidth]{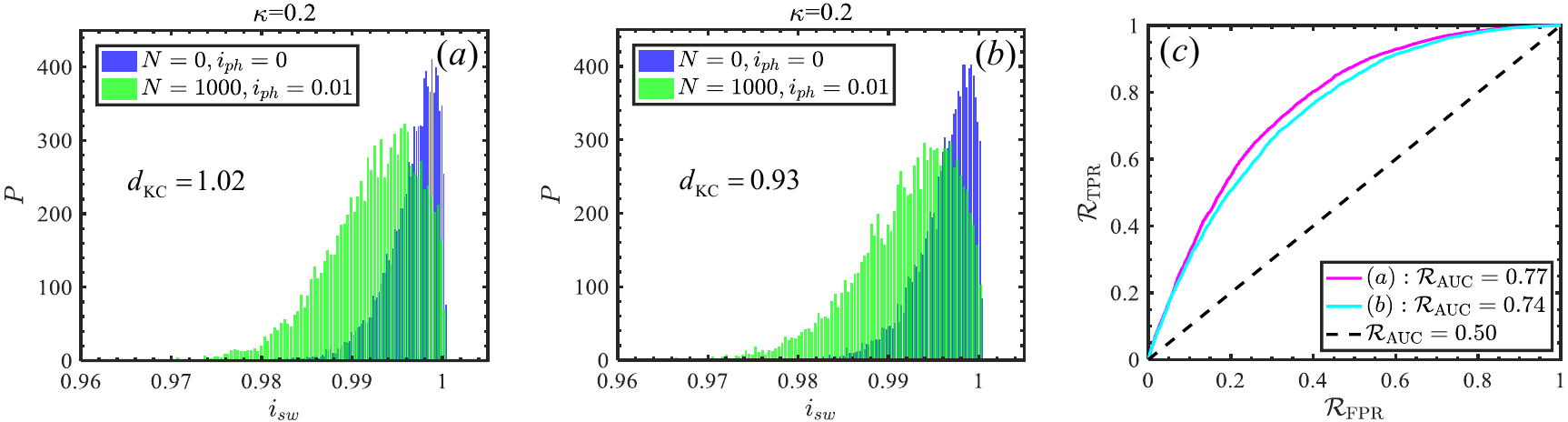}
\caption{The detectable simulations of an equilibrium JTD with $\kappa=0.2$, for a weak microwave pulsed signal with $i_{ph}=0.01$. The $d_{KC}$-indexes, shown respectively in (a) with $\varphi_0=0.1$ and (b) with $\varphi_0=0.2$, indicate that such a pulsed signal can be detected by using the equilibrium JTD, even its initial phase is different, while subfigure (c) indicates that such a detectability is significantly low and also insensitive to the initial phases. The parameters of the microwave pulsed signal are set as $\omega_{ph}=1$, $\tau_{ph}=1$, and $\tau_d=1/2v=1/2\kappa\beta=25000$, representing that the arrival time of the microwave pulse is half the duration of a single bias current sweep. The other parameters are set the same as in Fig.~\ref{FF2}.}\label{F12_1}
\end{figure}

Again, for the comparison, we first investigate the detectability of an equilibrium JTD for the pulsed microwave detection. For the pulsed signal with $N=1000$, we can see from Figs.~\ref{F12_1}(a-b) that, one can see the SCDs obtained by using the equilibrium JTD (with $\kappa=0.2$) are really single-peak shapes and insensitive to the initial phase $\varphi_0$. Figs.~\ref{F12_1}(c) shows that the values of $\mathcal{R}_{\rm AUC}$-index are manifestly low (just a little larger than the $\mathcal{R}^T_{\rm AUC}=0.7$), although their $d_{KC}$-indexes indicate that such a pulsed signal should be detectable. This implies that the ability of the equilibrium JTD is limited, even for the present relatively strong pulsed signal.  
While, as shown in Fig.~\ref{F10_2}(a, b), if the non-equilibrium JTD with $\kappa=5$ is utilized to detect such a pulsed signal, the changes of the SCD-fingerprints induced by the detected signal become more distinguishable. The vaules of the values of $\mathcal{R}_{\rm AUC}$-index can reach $0.81$ for $\varphi_0=0.1$ and $0.99$ for $\varphi_0=0.2$, respectively. These numerical results indicate that, for a common pulsed signal, the detection ability of the non-equilibrium JTD is manifestly stronger than that of its equilibrium counterpart. Therefore, given that the equilibrium JTD can be used for the detection of the microwave weak signal with a few photons~\cite{our1}, it is rationally expected that the non-equilibrium JTD can possess the ability for the detection of a microwave pulsed signal with just a single photon, and its achievable detection sensitivity can arrive at the desired single-photon level.  
\begin{figure}[htbp]
\includegraphics[width=1\linewidth]{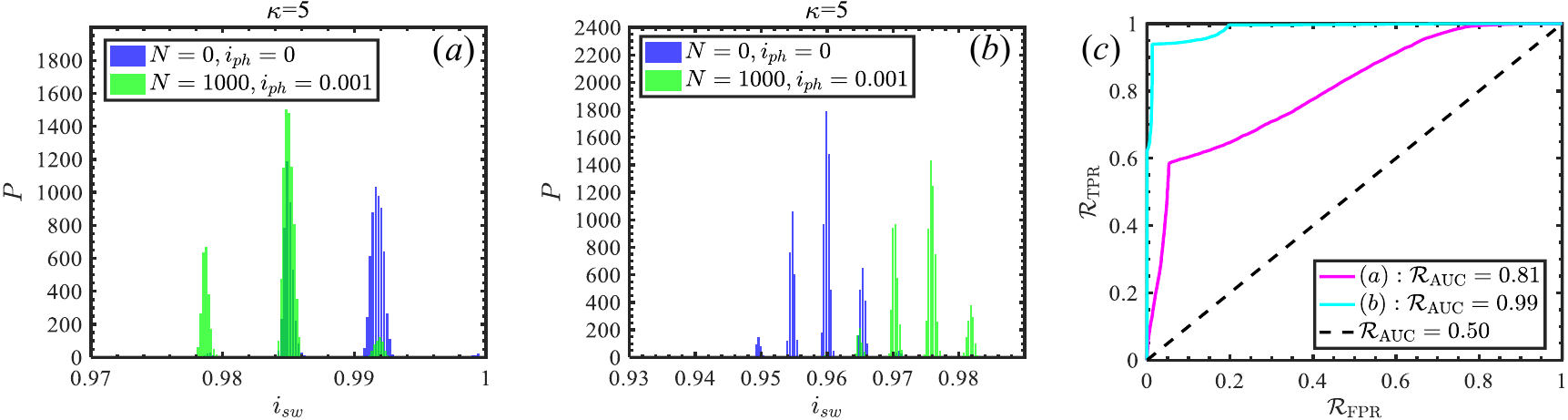}
\caption{The detectable simulations of a non-equilibrium JTD with $\kappa=5$ for the same weak microwave pulsed signal with $i_{ph}=0.001$; (a) for $\varphi_0=0.1$, (b) for $\varphi_0=0.2$. Subfigure (c) shows that its detectability is significantly higher compared with that of using the equilibrium JTD. Here, all the other relevant parameters are set the same as those in Fig.~\ref{F12_1}.} \label{F12_2}
\end{figure} 

\begin{figure}[htbp]
\includegraphics[width=1\linewidth]{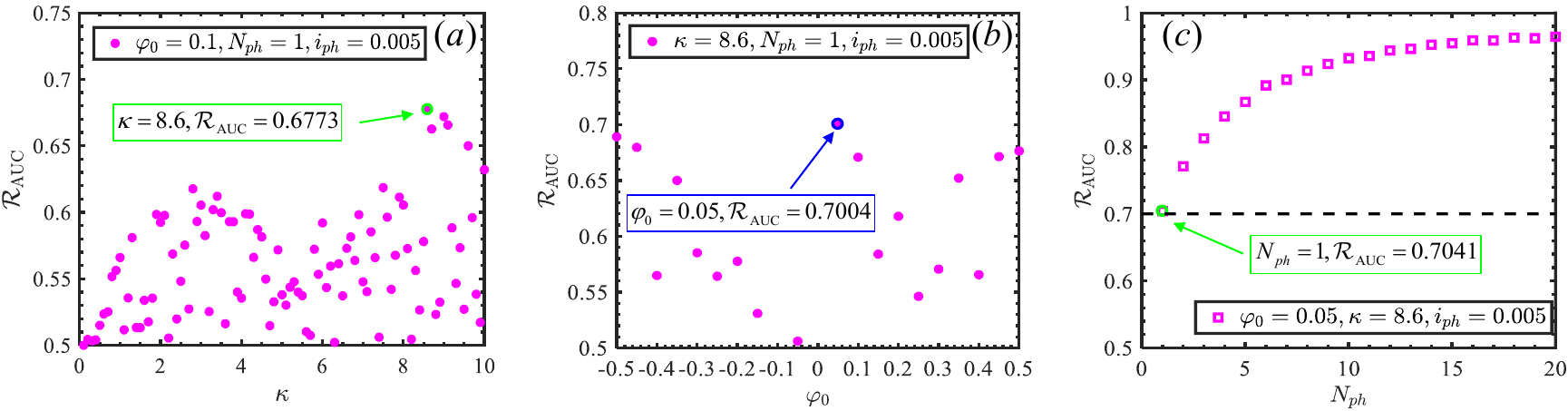}
\caption{Detectable simulations of a microwave single-photon pulsed signal by optimizing the operational parameters of the non-equilibrium JTD; (a) for the $\kappa$-parameter, (b) for the initial phase. Subfigure (c) clearly demonstrates that the non-equilibrium JTD with the optimized parameters can really be utilized to implement the desired microwave single-photon detection. Here, the parameters of a microwave single-photon pulsed signal are set as: $\omega_{ph}=1$, $\tau_{ph}=356$ (namely 5~ns), and $\tau_d=1/2v=1/2\kappa\beta$, indicating that the arrival time of the microwave pulse is half the duration of a single bias current sweep. The other parameters are set the same as in Fig.~\ref{FF2}.}\label{F13}
\end{figure}

Based on the above numerical results, Fig.~\ref{F13} presents the optimized operational parameters of the non-equilibrium JTD for the sensitive pulsed microwave signal detection. 
It is shown in Fig.~\ref{F13}(a, b) that the optimal sweep rate of the biased current is $\kappa=8.6$ and the optimal initial phase is $\varphi_0=0.05$. Accordingly, the achievable value of the $\mathcal{R}_{\rm AUC}$-index of the non-equilibrium JTD with the optimized operational parameters $\kappa=8.6$ and $\varphi_0=0.05$ is demonstrated in Fig.~\ref{F13}(c) for the different pulse intensities (i.e., the input microwave photon numbers). It is shown that, even for the pulsed microwave signal with the effectively current  $i_{ph}^{min}=I_{ph}/I_c\approx5.7~\text{nA}/0.975~\mu\text{A}\approx0.005$~\cite{our1,our3,Resonant} for about $5$ ns pulsed width~\cite{tang_storage_2015}, the value of $\mathcal{R}_{\rm AUC}$ of the JTD with the present operational parameters is still higher than the reliably detectable threshold, i.e, $\mathcal{R}_{\rm AUC}^{min}>\mathcal{R}_{\rm AUC}^T = 0.7$.

Therefore, by using a non-equilibrium JTD, with the optimized $\beta$-parameter and the operational parameters such as the sweep rate $\kappa$, initial phase $\varphi_0$, the detection of a single microwave-photon signal is experimentally possible, at least theoretically, whether such a signal is a continuous-wave or pulsed one. Probably, this is due to the fact that the achievable detection sensitivity of the non-equilibrium JTD proposed here is insensitive to the thermal noise, which is, however, one of the main obstacles to further improving the achievable detection sensitivities of the usual JTDs working in their equilibrium states.  
\subsection{Performance Metrics}
As a weak signal detection device that can be experimentally realized, it is essential to clarify various performance parameters when discussing its practicality. The main performance parameters of the non-equilibrium JTD proposed in the present work, for detecting the weak microwave signals at a single-photon level, typically include the detection bandwidth, dynamic range, and photon number resolution. In this subsection, we will conduct a numerical evaluation of these typical performance parameters, while the other parameters, such as the counting rate, dark counts, and the detection efficiency, will be evaluated and calibrated in future experimental studies through the actual measured data of the device.

\begin{figure}[htbp]
\includegraphics[width=1\linewidth]{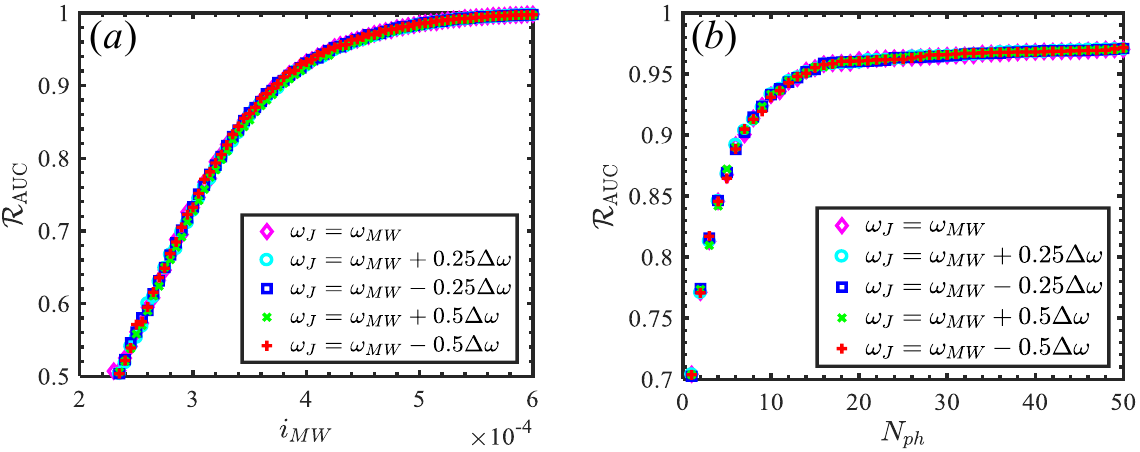}
\caption{The numerically simulated detection bandwidths and dynamic ranges of the non-equilibrium JTD with the optimized parameters; (a) for the continuous-wave single microwave-photon signal, and all the parameters, except the signal frequency, match those in Fig.~\ref{F11}(c); (b) for the pulsed single microwave-photon signal. Here, all the parameters, except the signal frequency, match those in Fig.~\ref{F13}(c).}\label{F14}
\end{figure}

First, it is believed that the detection bandwidth of the proposed detector is sufficiently large, although the above discussion on its detectability in Sec.~\ref{sec:3B} is based on the signal with the frequency being equivalent to the plasma frequency $\omega_J$ of the JTD. In fact, we argue that the signals with the frequencies being within the 3-dB bandwidth centered at $\omega_J$ remain detectable. This is because, with the quality factor (Q-factor) formula for the underdamped JTDs~\cite{PhysRevApplied.22.024015}
\begin{equation}
\begin{aligned}
Q=\omega_J RC\,,
\end{aligned}
\end{equation}
the normalized effective operational bandwidth $\Delta\omega$ (relative to $\omega_J$) of the detector can be estimated as
\begin{equation}
\begin{aligned}
\Delta\omega=\frac{\omega_J}{Q\omega_J}=\beta=10^{-4}.
\end{aligned}
\end{equation}
In Figs.~\ref{F14}(a) and (b), we numerically simulate the detectabilities for the continuous-wave and pulsed single-photon signals, respectively, within this bandwidth. The parameter settings are identical to those in Fig.~\ref{F11}(c) and Fig.~\ref{F13}(c), with the only difference being the frequency of the input microwave signal. As shown in Fig.~\ref{F14}(a), the $\mathcal{R}_{\rm AUC}$-index values are almost unchanged for the continuous-wave microwave signals with the frequencies within the range of $[\omega_J-\Delta\omega/2,\omega_J+\Delta\omega/2]$, indicating no variation in detectability and thus no change in the achievable detection sensitivity across this bandwidth. Fig.~\ref{F14}(b) clearly demonstrates that, for the pulsed microwave signals within this bandwidth, all $\mathcal{R}_{\rm AUC}$-index values exceed the threshold $\mathcal{R}_{\rm AUC}^T=0.7$, confirming that the detection of any pulsed single microwave-photon signals is still feasible, if its frequency is really within this range. 

Second, for the JTD proposed here for the single microwave-photon detection, the dynamic range can be defined as the region within which its response to the intensity of an input weak microwave signal is approximately linear~\cite{9769720}. In other words, within this dynamic range, there exists a one-to-one correspondence between the $\mathcal{R}_{\rm AUC}$-index value and the microwave signal strength. This means that the JTD can not only detect microwave weak signals within this intensity range but also distinguish their relative strengths. When the microwave signal strength increases to the point where $\mathcal{R}_{\rm AUC}$ no longer scales linearly with it, the JTD can no longer reliably distinguish between different signal strengths. Therefore, for a JTD functioning as a single microwave-photon detector, the upper bound of its dynamic range can defined as the the microwave signal strength at which $\mathcal{R}_{\rm AUC}$-index ceases to the linearity for the increasing signal strength, while the lower bound can be similarly defined as the signal strength at which the JTD's $\mathcal{R}_{\rm AUC}$-index just reaches the threshold $\mathcal{R}_{\rm AUC}^T=0.7$. As shown in Fig.~\ref{F14}(b), even for a single microwave-photon signal, the JTD's $\mathcal{R}_{\rm AUC}$-index already exceeds this threshold. Thus, for a JTD applicable to the single microwave-photon detection, the lower limit of its dynamic range is practically zero. For example, the JTD corresponding to the dotted curve in Fig.~\ref{F14}(b) exhibits a dynamic range for the weak pulsed microwave signals with the photon numbers being in the interval $[1,15]$, which implies that its maximal resolvable number of the microwave photons is $N_{ph}^{max}=15$ for the present detector. 

Consequently, based on the above discussions on the dynamic range of the proposed detector, the non-equilibrium JTDs for the single microwave-photon detection inherently possess the photon-number resolution, making it a photon-number-resolving single microwave-photon detector. Clearly, the performance metrics of the JTD depend basically on its working state, either in the usual equilibrium or the present non-equilibrium state. This can be controlled by adjusting the physical and operational parameters of the detector. Therefore, one advantage of JTDs, compared to other weak microwave detectors, is their adjustable dynamic ranges, making them applicable for detecting weak microwave signals of different intensities.
\section{conclusion and discussion}
In summary, by numerically solving the phase dynamical equation of the JTD with the different initial phases and the sweep rates of the biased current, we find that the JTD can be operated in either the conventional equilibrium state or the non-equilibrium state, depending on the JTD being driven either adiabatically or non-adiabatically. As its phase dynamics are sensitive to the existing thermal noise, the usual equilibrium JTD can only be utilized to detect the microwave weak signal approaching the single-photon energy limit. Differently, as the phase dynamics is insensitive to the existing thermal noise, the non-equilibrium JTD has the ability to detect weak microwave signals with the detection sensitivity of a single microwave-photon. 
Using the binary detection criterion in statistics, we demonstrated numerically the detectability of a single microwave-photon signal by using the non-equilibrium JTD, no matter whether the input signal is a continuous-wave or a pulsed one. In particular, we demonstrated that such a non-equilibrium JTD can naturally possess the ability for the photon-number-resolvable detection of the microwave weak signal. 

Certainly, since the detection approach proposed here requires distinguishing two statistical distributions of the measured switching currents obtained by multiple measurements, the present non-equilibrium JTD cannot be utilized to implement the single-shot detection of a microwave single-photon signal. Probably, various developing large-data machine learning methods are useful for improving the distinguished velocity of the SCDs for the present non-equilibrium JTD, which is feasible with the current technology. 
\begin{acknowledgments}
This work is partially supported by the National Key Research and Development Program of China (NKRDC) under Grant No. 2021YFA0718803, and the National Natural Science Foundation of China (NSFC) under Grant No. 11974290.
\end{acknowledgments}
\appendix
\section{\label{sec:appendix.A}
Adiabatic and non-adiabatic evolution conditions in nonlinear dynamical systems}
In Sec.~II. B, we numerically demonstrated that the sweep rate of the bias current determines whether the JTD is driven adiabatically or non-adiabatically, thereby governing whether it is operated at the equilibrium or non-equilibrium state. In this appendix, we provide a detailed analysis of the adiabatic driving condition of the JTD.

Physically, the dynamics for a driven system are often described by multiple parameters. Among these, some vary on very short time scales and thus they rapidly approach certain steady-state values; these parameters are usually referred to as the fast variables. While the others are referred to as the slow variables, which change so slowly that they can be considered as invariant over the time scales of the fast dynamics. The adiabatic approximation, therefore, refers to the scenario in which, during the gradual evolution of the slow variables, the fast variables instantaneously adjust to their equilibrium states. As a result, the dynamical influence of the fast variables can be neglected throughout the slow variation process~\cite{griffiths}.

To characterize the dynamical behavior of these two types of variables, the characteristic time scales, $\tau_{fast}$ and $\tau_{slow}$, are introduced to represent the fast and slow variables, respectively. These time scales should satisfy the following relation~\cite{griffiths}:
\begin{equation}
\begin{aligned}
\tau_{fast}<<\tau_{slow}.
\end{aligned}
\end{equation}
As a consequence, the relative timescale parameter can be defined as
\begin{equation}
\begin{aligned}
\epsilon=\frac{\tau_{fast}}{\tau_{slow}},
\end{aligned}\label{eq:Nonadiabatic}
\end{equation}
therefore, the condition $\epsilon<<1$ is then referred usually as the adiabatic condition.

Typically, the dynamics of a driven damped pendulum under a slowly varying external force can be described by the following equation:
\begin{equation}
\begin{aligned}
ml^2\ddot{\theta}+b\dot{\theta}+mgl\sin(\theta)=F(t)\,,
\end{aligned}
\end{equation}
where $m$, $l$, $\theta$, and $b$ represent the mass of the pendulum bob, the length of the pendulum, the angular displacement, and the damping coefficient, respectively. The external driving force, which varies with time, is given by $F(t)=F_0t$ with $F_0$ being constant. Obviously, the fast variable of this dynamical system is the angular velocity $\dot{\theta}$, as the damping term $b\dot{\theta}$ causes the system to rapidly converge to damped oscillations around the equilibrium position. The slow variable of the system is the external driving force $F(t)$. When the system is dominated by the damping term, the effects of the external force and gravity can be neglected, reducing the behavior to that of such a simply damped oscillation without driving, i.e.,
\begin{equation}
\begin{aligned}
ml^2\ddot{\theta}+b\dot{\theta}\approx0.
\end{aligned}
\end{equation}
The formal solution of this equation can be expressed as:
$\dot{\theta}\propto e^{-bt/(ml^2)}$,
yielding the characteristic time scale of the fast variable reads:
\begin{equation}
\begin{aligned}
\tau_{fast}=\frac{ml^2}{b}\,.
\end{aligned}
\end{equation}
As the change rate of the external force is $\dot{F}$, and its characteristic time scale can be defined as the time over which the force changes significantly, i.e.,
\begin{equation}
\begin{aligned}
\tau_{slow}=\frac{F}{\dot{F}}=t\,,
\end{aligned}
\end{equation}
the condition for the pendulum oscillation to be adiabatic is thus:
\begin{equation}
\begin{aligned}
\epsilon=\frac{\tau_{fast}}{\tau_{slow}}=\frac{ml^2}{b}<<t.
\end{aligned}
\end{equation}

Similarly, for the phase dynamics of the present JTD without the noise, as discussed in the text, the equation of motion reads
\begin{equation}
\begin{aligned}
\frac{{\rm d}^2\varphi}{{\rm d}\tau^2}+\beta\frac{{\rm d}\varphi}{{\rm d}\tau}+\sin(\varphi)=v\tau\,,
\end{aligned}\label{eq:A8}
\end{equation}
which is formally identical to the dynamical equation of the driven damped pendulum described above. Therefore, the characteristic time scale of the fast variable in the phase dynamics can be taken as
\begin{equation}
\begin{aligned}
\tau_{fast}=\frac{1}{\beta},\,
\end{aligned}
\end{equation}
while the time scale of the slow observable is
\begin{equation}
\begin{aligned}
\tau_{slow}=\frac{i_b}{\dot{i}_b}=\tau.
\end{aligned}
\end{equation}
Thus, the adiabatic condition for the phase evolution governed by equation~\eqref{eq:A8} is
\begin{equation}
\begin{aligned}
\epsilon=\frac{\tau_{fast}}{\tau_{slow}}=\frac{1}{\beta\tau}=\frac{v}{\beta}<<1,
\end{aligned}\label{eq:A11}
\end{equation}
as $v\tau=1$ for the present JTD. Hence, the duration for which the biased current is applied must satisfy the condition: $\tau\leq1/v$. Therefore, $v/\beta=1$ refers to the critical condition distinguishing the adiabatic and the non-adiabatic evolution of the phase dynamics for the JTD. This condition actually served as the physical basis for defining whether the JTD is operating in an equilibrium or a non-equilibrium state. 

Consequently, if the adiabatic condition defined by Eq.~\eqref{eq:A11} is satisfied, the phase dynamical evolution for the JTD is dominated by the damping term (originated basically from the thermal noise) and thus it is insensitive to the initial phase. However, when the adiabatic condition, i.e., Eq.~\eqref{eq:A11} is violated, the phase dynamics for the JTD should be governed by the time-dependent bias current, rendering the evolution highly sensitive to the initial phase. In this case, the JTD is referred to as being operated in a non-equilibrium state.
\section{\label{sec:appendix.B} Numerical demonstration of effective thermal noise suppression in a non-equilibrium JTD device}
In Sec.~II.~C, we argued that the phase dynamics for a non-equilibrium JTD being insensitive to the thermal noise, yielding its achievable detection sensitivity can be greatly enhanced. In this appendix, we numerically verified this argument. 

\begin{figure}[htbp]
\includegraphics[width=0.8\linewidth]{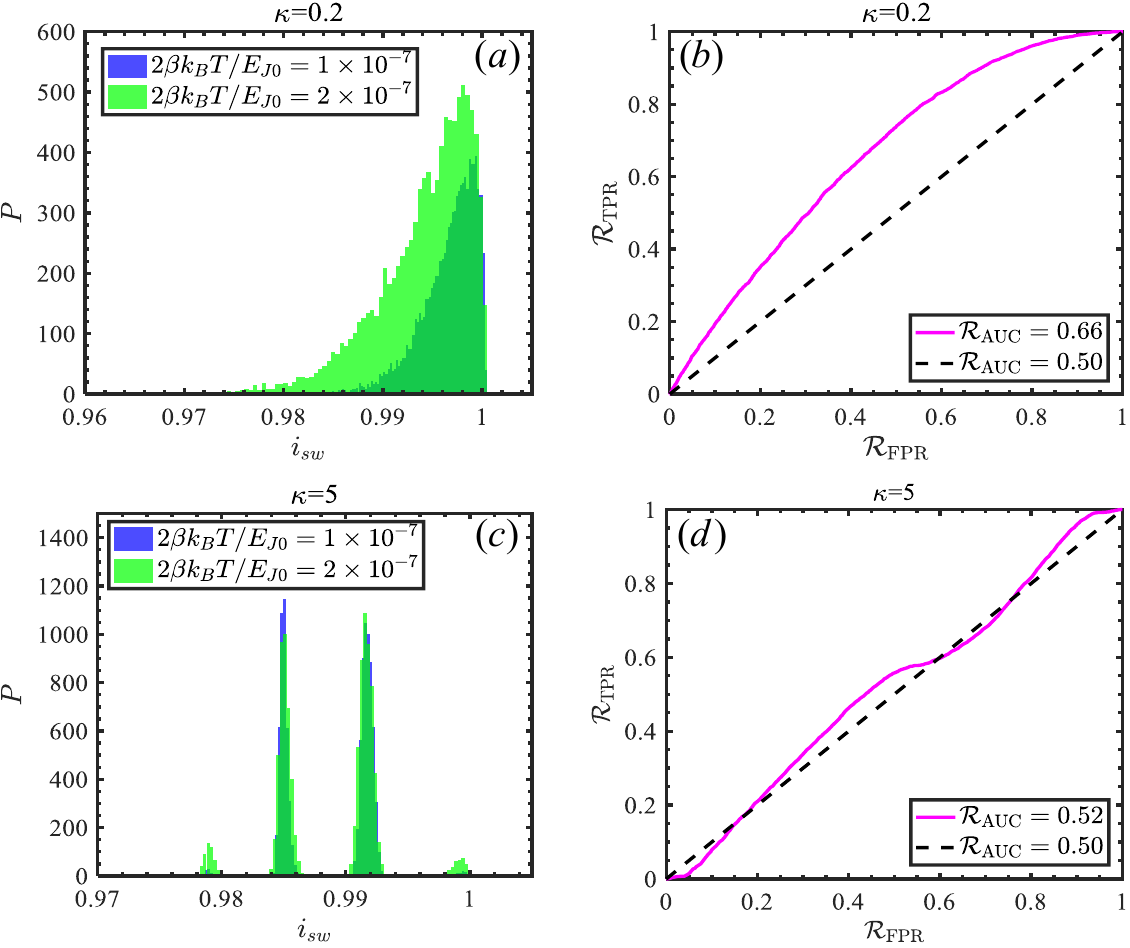}
\caption{Numerically verify the thermal noise sensitivity of the SCDs for (a-b) the equilibrium JTD with $\kappa=0.2$, and (c-d) the non-equilibrium JTD with $\kappa=5$. Here, the relevant parameters are set as; $\varphi_0=0.1$, $\dot{\varphi_0}=0$, $\beta=10^{-4}$.}\label{F15}
\end{figure}

For the comparison, Figs.~\ref{F15}(a) and (c) present the SCDs of the JTD under equilibrium ($\kappa=0.2$) and non-equilibrium ($\kappa=5$) states, respectively, at different operating temperatures (corresponding to varying levels of thermal noise current). 
It is seen from Fig.~\ref{F15}(a) that the SCDs of the equilibrium JTD at different temperatures all exhibit unimodal structures. Thus, the distinguishability of the SCDs, i.e., $\mathcal{R}_{\rm AUC}=0.66>0.5$ can be obtained for the ROC curve shown in Fig.~\ref{F15}(b). This indicates clearly that the thermal noise significantly limits the achievable detection sensitivity of the equilibrium JTD. In contrast, for $\kappa=5$, i.e., the JTD is operated in its non-equilibrium state, Fig.~\ref{F15}(c) reveals that the SCDs under different thermal noise intensities are no longer unimodal. Moreover, for the ROC curve shown in Fig.~\ref{F15}(d), we have $\mathcal{R}_{\rm AUC}=0.52$. This indicates that the variation of thermal noise is almost indistinguishable in the SCD for the non-equilibrium JTD. 

Therefore, its phase dynamics are really insensitive to the thermal noise. As a consequence, the achievable detection sensitivity of the non-equilibrium JTD can surpass that of the equilibrium JTD, which is limited by a few microwave photon levels. This indicates that, when the JTD is employed for the weak microwave signal detection, the non-equilibrium JTD, instead of the usual equilibrium JTD, can overcome the imposed by the thermal noise and thus can achieve a single microwave-photon detection.
\section{\label{sec:appendix.C} A feasible approach to accurately modulate the initial phase of the JTD}

The numerical results are demonstrated in Secs.~II.~B and II.~C indicated that the phase dynamics for the non-equilibrium JTD, with the large sweep rate of the biased current, are strongly sensitive to the initial phase, yielding the unique ``fingerprint" in its SCD. Therefore, to implement the desired single-photon detection by distinguishing these ``fingerprint" changes in the SCDs, we need to precisely modulate the initial phase of the JTD. Indeed, this is feasible. 

\begin{figure}[htbp]
\includegraphics[width=0.7\linewidth]{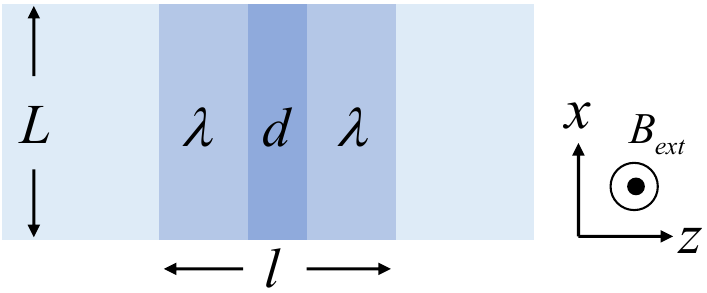}
\caption{A simple approach to modulate the initial phase of the JTD by using an external magnetic field  $B_{ext}$ generated by an off-line current, which is perpendicular and directed into the plane of the paper. Here, $L$ denotes the width of the JTD, $d$ is the thickness of the insulating layer, and $\lambda$ represents the London penetration depth of the superconductors on both sides of the insulating layer. Thus, $l=d+2\lambda$ corresponds to the effective length over which the external magnetic field acts on the junction region.}\label{FF6}
\end{figure}

Actually, as shown in Fig.~\ref{FF6}, the initial phase of the JTD can be modulated effectively by using the external magnetic flux, which is generated by an off-line current. Here, according to the Meissner effect, the magnetic field used to modulate the initial phase of the JTD can be expressed as 
\begin{equation}
B_{ext}(\tau)=
\begin{cases}
B_0, & \tau_0\leq \tau\leq \tau_0+\tau_{ext}\,, \\
0, & \text{other}\,.
\end{cases}
\end{equation}
Here, $B_0$, $\tau_0$, and $\tau_{ext}$ represent the magnetic flux density, the initial time of the field generation, and the duration of the field, respectively. Therefore, under the modulation of this external magnetic field, the initial phase of the JTD can be expressed as~\cite{JJmagnetic}
\begin{equation}
\begin{aligned}
\varphi_0=\widetilde{\varphi}_0+\frac{2elLB_0}{\hbar}\,,
\end{aligned}\label{eq:magnetic}
\end{equation}
where $\widetilde{\varphi}_0$ represents the initial phase in the absence of external magnetic field modulation. For the typical JTD parameters: $d\approx1.2$~nm, $L\approx1.5 \rm~\mu m$~\cite{our2,our3}, and taking the London penetration depth of the superconducting aluminum as~$16$~nm~\cite{AL_LD}, we have $\varphi_0-\widetilde{\varphi}_0\approx 151 B_0$. This indicates that an external magnetic field with a flux density of only several tens of Gauss is sufficient to effectively modulate the initial phase of the JTD.

\bibliography{Ref}
\end{document}